\documentclass[preprint, prb, superscriptaddress]{revtex4-2}
\usepackage[utf8]{inputenc}
\usepackage{braket}

\usepackage{graphicx}
\usepackage{amsmath}

\usepackage[colorlinks = true,
            linkcolor = blue,
            urlcolor  = blue,
            citecolor = blue,
            anchorcolor = blue]{hyperref}
            
\usepackage{tikz,xcolor,hyperref}

\definecolor{lime}{HTML}{A6CE39}
\DeclareRobustCommand{\orcidicon}{
	\begin{tikzpicture}
	\draw[lime, fill=lime] (0,0) 
	circle [radius=0.16] 
	node[white] {{\fontfamily{qag}\selectfont \tiny ID}};
	\draw[white, fill=white] (-0.0625,0.095) 
	circle [radius=0.007];
	\end{tikzpicture}
	\hspace{-2mm}
}
\foreach \x in {A, ..., Z}{%
	\expandafter\xdef\csname orcid\x\endcsname{\noexpand\href{https://orcid.org/\csname orcidauthor\x\endcsname}{\noexpand\orcidicon}}
}

\newcommand{\vxl}{v_{x \lambda} }

\newcommand{\fl}{f_{\lambda}}
\newcommand{\fll}{f_{\lambda, l}}
\newcommand{\flz}{f_{\lambda, 0}}
\newcommand{\flo}{f_{\lambda, 1}}
\newcommand{\flp}{f_{\lambda'}}
\newcommand{\flpz}{f_{\lambda',0}}

\newcommand{\flpl}{f_{\lambda',l}}

\newcommand{\phil}{\Phi_{\lambda}}

\newcommand{\philo}{\Phi_{\lambda,1}}
\newcommand{\philpo}{\Phi_{\lambda',1}}

\newcommand{\ezl}{e_{0,\lambda}}
\newcommand{\ezlp}{e_{0,\lambda'}}

\newcommand{\xbar}{\overline{x}}
\newcommand{\Cllp}{C_{\lambda \lambda'} }
\newcommand{\M}{M_{\lambda \lambda'}}

\usepackage{etoolbox}
\usepackage{caption}
\usepackage{subcaption}
\usepackage[symbol]{footmisc}

\makeatletter

\patchcmd{\frontmatter@abstract@produce}
  {\vskip200\p@\@plus1fil
   \penalty-200\relax
   \vskip-200\p@\@plus-1fil}
  {}
  {}
  {}
\makeatother

\begin{document}
\title{Quasiballistic electron transport in cryogenic SiGe HBTs studied using an exact, semi-analytic solution to the Boltzmann equation}

\author{Nachiket R. Naik\orcidA{}}
\affiliation{Division of Engineering and Applied Science, California Institute of Technology, Pasadena, CA 91125, USA}
\author{Austin J. Minnich \footnote[1]{Corresponding author (aminnich@caltech.edu)} \orcidB{}} 
\affiliation{Division of Engineering and Applied Science, California Institute of Technology, Pasadena, CA 91125, USA}
\email{aminnich@caltech.edu}
\date{\today}

\begin{abstract}

Silicon-germanium heterojunction bipolar transistors (HBTs) are of interest as low-noise microwave amplifiers due to their competitive noise performance and low cost relative to III-V devices. The fundamental noise performance limits of HBTs are thus of interest, and several studies report that quasiballistic electron transport across the base is a mechanism leading to cryogenic non-ideal IV characteristics that affects these limits. However, this conclusion has not been rigorously tested against theoretical predictions because prior studies modeled electron transport with  empirical approaches or approximate solutions of the Boltzmann equation. Here, we study non-diffusive transport in narrow-base SiGe HBTs  using an exact, semi-analytic solution of the Boltzmann equation based on an asymptotic expansion approach. We find that the computed transport characteristics  are inconsistent with experiment, implying that quasiballistic electron transport is unlikely to be the origin of cryogenic non-ideal IV characteristics. Our work helps to identify the mechanisms governing the lower limits of the microwave noise figure of cryogenic HBT amplifiers.  

\end{abstract}

\maketitle

\clearpage

\section{Introduction}

Silicon-germanium heterojunction bipolar transistors (HBTs) are widely used in microwave applications such as  radar and communication systems \cite{rucker2012halfthz, Chevalier2017} and show potential in space science \cite{bardin2013high} and imaging applications \cite{Yuan2009,cressler2021new, frounchi2020sige} owing to their competitive microwave performance combined with ease of integration with the CMOS process, high yield, and low cost relative to III-V technologies \cite{cressler2005potential,cressler2003silicon}. SiGe HBTs are now approaching the THz domain due to fabrication process advancements that enable the continued scaling of key parameters such as base width and resistance \cite{Chevalier2017, Heinemann2017}. Moreover, at cryogenic temperatures they have demonstrated DC gains exceeding 45 dB, noise temperatures $\lesssim  5$ K at 3 GHz with bandwidths of 2-3 GHz \cite{Bardin2017,Bardin2009}, lower $1/f$ noise relative to III-V devices \cite{van1998low}, and power consumption in the hundreds of microwatts \cite{Montazeri2016}. These competitive figures of merit make SiGe HBTs of interest for low-noise amplifiers and cryo-CMOS circuits \cite{Wong2020, Patra2018} in radio astronomy and quantum computing applications \cite{Weinreb2007, Montazeri2016} and in oscillators with low phase noise \cite{van1998low, plana1997, Niu2005}.

Following the first reports of SiGe HBTs grown by molecular beam epitaxy \cite{Patton1988a, gruhle1992mbe}, studies of the cryogenic performance of HBTs were performed in the early 1990s \cite{Patton1988a, Patton1990}. Subsequent work  focused on understanding and optimizing base doping and Ge grading profiles for cryogenic performance \cite{Cressler1993}. While these works demonstrated the enhanced collector current and transconductance expected at cryogenic temperatures, later studies reported marked deviations from the DC current-voltage characteristics predicted from drift-diffusion theory \cite{Joseph1995, Richey1996}. Specifically, it was observed that below $\sim 77$ K the collector $(I_C)$ and base current $(I_B)$ exceeded the predicted values at a given temperature and bias and were independent of temperature, with the transconductance $g_m$ saturating instead of increasing as $T^{-1}$. These non-ideal $IV$ characteristics limit the gain, cutoff frequency, and ultimately the microwave noise figure.

The origin of this behavior has been attributed to several mechanisms, including trap-assisted tunneling at base-emitter voltage ($V_{BE}$) well below the built-in potential; \cite{Joseph1995} and non-equilibrium carrier transport, in which electrons quasiballistically traverse the base, at biases comparable to the built-in potential \cite{Richey1996}. The latter effect has been phenomenologically described using an elevated electron temperature that is taken as the effective temperature for thermionic emission at the base-emitter junction \cite{Bardin2009, Richey1996, Bardin2017}. As base widths were scaled further down to $\sim 20$ nm in recent years, direct tunneling of electrons was reported to contribute to the collector current at biases approaching the built-in potential \cite{Rucker2017, Davidovic2017,Ying2018}.

There are several discrepancies with these explanations, however. First, non-ideal collector currents have been reported in first generation devices with base widths on the order of $100$ nm \cite{Ying2018}, for which direct tunneling is unlikely.  Further, at base doping levels above $\sim 3\times 10^{18}$ cm$^{-3}$ common in modern devices, \cite{Ying2018,Rucker2017} ionized impurity scattering is expected to dominate. Therefore, quasiballistic transport is not expected to be markedly more pronounced at cryogenic temperatures relative to room temperature as evidenced by the weak temperature dependence of the minority carrier mobility \cite{Swirhun1988,Klaassenn1992,Rieh2000a,Bardin2009}. Finally, while electrons may be heated as they traverse the base by the built-in field, the collector current and transconductance depend on the electron temperature at the base-emitter junction prior to transport across the base and thus cannot be affected by the built-in field. 

A quantitative, non-empirical description of the quasiballistic transport process would allow a more thorough test of whether quasiballistic transport is a possible origin of the cryogenic DC non-idealities. Steady particle transport across a slab of thickness comparable to the MFPs of the particles is described by the Boltzmann equation \cite{Lundstrom2000fundamentals,chen2005nanoscale}. The slab problem has been extensively investigated for radiative \cite{chen2005nanoscale, chandrasekhar1960radiative, modest2013radiative}, neutron \cite{spanier1969monte}, phonon \cite{chen2005nanoscale, Hua2015}, and electron transport \cite{grinberg1992diffusion,Stettler1994, Tanaka, ozaydin1996non} as well as for rarefied gases \cite{cercignani1963flow, cercignani2000rarefied, Manela2008}. Although analytical solutions under the 'one-speed' or 'gray' approximations are possible,\cite{grinberg1992diffusion, modest2013radiative} solutions considering energy-dependent relaxation times are less tractable analytically. Alternative numerical approaches are computationally expensive owing to the need to discretize both the spatial and reciprocal space coordinates \cite{grinberg1992diffusion,Stettler1994,Lundstrom2018}. Several works have reported asymptotic series expansion methods, based on the original theory by Grad \cite{grad1963asymptotic}, for rarefied gas dynamics \cite{sone1969asymptotic} and phonon transport \cite{Peraud2016a}. These methods could enable the efficient solution of the present problem of electronic transport in a narrow base, but they  have yet to be adapted for electronic transport.

Here, we report a study of quasiballistic electron transport across a narrow base using an exact, semi-analytic asymptotic expansion approach to solve the Boltzmann equation. We show that quasiballistic transport leads to decreased collector current for a given base-emitter voltage and does not alter the transconductance. Both of these findings contradict experimental observations, indicating that quasiballistic transport is unlikely to be the origin of cryogenic non-idealities in SiGe HBTs. Instead, we hypothesize that the origin of these non-idealities could arise from local inhomogeneities in built-in potential which are readily observed in Schottky diodes \cite{Tung1991nisi,Tung1992} and have been postulated to exist in SiGe HBTs \cite{vonHaartman2002rts}. Our work advances the understanding of cryogenic transport processes in HBTs that limit gain, transconductance, and thus ultimately microwave noise performance.

\section{Theory}

We begin by describing the semi-analytical asymptotic approach used to solve the one-dimensional Boltzmann equation describing electron transport across the base. Focusing on the DC characteristics, we assume steady transport across a base of width $L$ with  prescribed forward and reverse going electron distribution functions at the left and right boundaries, respectively. Then, the Boltzmann equation for the electronic distribution function $\fl(x)$ is given by:

\begin{eqnarray}
\label{eq:BE1}
    \frac{\partial \fl}{\partial \xbar} &=& \frac{L}{\vxl}\sum_{\lambda'} \Cllp \flp = \sum_{\lambda'} \M \flp
\end{eqnarray}
where $\Cllp$ is the  collision matrix, $\vxl$ is the group velocity for electronic state $\lambda$, $\M$ is a  dimensionless  matrix defined as above, $x$ is the spatial coordinate, $ \xbar \equiv x/L$, and $\fl$ is the desired distribution function normalized so that $V^{-1} \sum_{\lambda} \fl (x) = n(x) $, where $V$ is a normalizing volume. Macroscopic quantities like electric current are computed using the appropriate Brillouin zone sum~\cite{chen2005nanoscale}.

In this work, we will solve this equation using  an exact, semi-analytic asymptotic expansion method originally reported for rarefied gases \cite{sone1969asymptotic} and recently adapted for phonons \cite{Peraud2016a}. The theory for phonons is given in detail in Ref.~\cite{Peraud2016a} for an isotropic crystal under the relaxation time approximation. Here, we generalize this theory to incorporate the full electron collision matrix and arbitrary crystals and extend its applicability beyond linear deviations from a reference distribution.

In the asymptotic expansion approach,  $\fl$ and associated observables are written as series expansions with the average Knudsen number $\epsilon \equiv \braket{\Lambda}/L$ as the expansion parameter, where the average mean free path $\braket{\Lambda} \equiv \sum_{\lambda_k} \vxl \tau_{\lambda} f_{\lambda}^{eq} / n_{eq} $. Here $f_{\lambda}^{eq}$ and $n_{eq}$ are the reference equilibrium distribution function and electron density at zero base-emitter voltage, and $\tau_{\lambda}$ is the wave vector dependent relaxation time. As an example, $\fl$ is written as:
\begin{eqnarray}
    \fl &=& \sum_l \epsilon^l \fll %\\
\end{eqnarray}
where $\fll$ is the distribution function for order $l$. Substituting this expansion into Eq.~\ref{eq:BE1}, we get
\begin{equation}\label{eq:BE2}
    \sum_l \epsilon^l \frac{\partial \fll}{\partial x} = \sum_l \epsilon^l \sum_{\lambda'} \M \flpl
\end{equation}
Since $\M$ scales with $\epsilon^{-1}$, the left and right side of Eq.~\ref{eq:BE2} are offset by one order of $\epsilon$. Equating terms of the lowest order $l=0$ we find that the zeroth-order solution satisfies
\begin{equation}
  \sum_{\lambda'} \M \flpz(x) = 0 = g_0(x) \sum_{\lambda'} \M \ezlp 
\end{equation}
where we have split the solution into a pure $x-$dependent term $g_0(x)$ and wave vector dependent term $\ezl$, i.e. $\flz(x) = \ezl g_0(x)$. We note here that the null eigenvector $\ezl$ can be any Boltzmann distribution function, but for this work we choose $\ezl = f_{\lambda}^{eq}$ corresponding to the Boltzmann distribution for the case where no bias is applied.

We now consider the first-order terms. From Eq.~\ref{eq:BE2}, the solution can be written using the zeroth-order solution:

\begin{eqnarray}\label{eq:fbulk1}
    \flo(x) = \sum_{\lambda'} \epsilon^{-1} \M^{-1} \left( \frac{\partial \flpz}{\partial \xbar} \right) + g_1(x)\ezl
\end{eqnarray}

This solution is also determined only up to an $x$-dependent multiple of the null eigenvector, denoted $g_1(x)$, which will be obtained after deriving the governing equation for $x$-dependence. The higher-order solutions proceed similarly.

The $x$-dependence of the zeroth-order solution can be derived by enforcing current continuity for the electric current density $J_x$ in the $x$-direction for a 1D steady slab:

\begin{equation}\label{eq:dJdx}
    \frac{\partial J_{x}}{\partial \xbar} = 0 = V^{-1} \frac{\partial}{\partial \xbar} \sum_{\lambda} q \vxl \sum_l \epsilon^{l} \fll
\end{equation}
which must be satisfied at each order of the expansion. Substituting from Eq.~\ref{eq:fbulk1} for the first-order term, we find that

\begin{equation}
    \sum_{\lambda'} \vxl \epsilon \frac{\partial}{\partial \xbar}  \left( \epsilon^{-1} \M^{-1} \frac{\partial \flpz}{\partial \xbar} + g_1(x) \ezl\right) = 0
\end{equation}
 
The second term in the parentheses vanishes since $\vxl$ is odd in $\lambda$ whereas $\ezl$ is even. Therefore, we find
 
 \begin{equation}
     \frac{\partial^2 \flz(x)}{\partial \xbar^2} = 0
 \end{equation}

Applying $V^{-1} \sum_{\lambda} \flz(x) = n_0(x)$ to the above equation, we see that the zeroth-order electron density $n_0(x)$ satisfies the diffusion equation. Similarly, it can be shown that higher-order terms of $\fll(x)$ also satisfy the diffusion equation (see Appendix A of Ref.~\cite{Peraud2016a}). 

To solve these equations, the boundary conditions at each order must be specified. The boundary conditions at zeroth order are simply the prescribed isotropic, hemispherical electron distribution functions at the edges of the slab specified by $n_L$ and $n_R$. Note that in general, the actual carrier density at the edges of the slab will differ from these values owing to quasiballistic transport. Thus, the solution at zeroth order is just the diffusion equation solution. Note that the boundary conditions of the original problem are completely satisfied at zeroth order.

We now discuss the boundary conditions at higher orders. Because the zeroth order solution completely satisfies the boundary conditions of the original problem, to enforce the boundary conditions at higher orders we must introduce a boundary correction term $\phil(x)$ that satisfies the Boltzmann equation in the boundary region. This boundary correction must exactly cancel the contribution from $\fl(x)$ at the boundaries and vanish in the bulk. 

The Boltzmann equation for the boundary solution  at first-order is:

\begin{equation}\label{eq:phi1}
    \frac{\partial \philo (\eta)}{\partial \eta} = \sum_{\lambda'} \epsilon \M \philpo(\eta)
\end{equation}

where we introduce the stretched boundary-region coordinate $\eta \equiv x /\epsilon$. Focusing on the left boundary at $\xbar = 0$, the condition enforced on this first-order boundary term is that it must cancel the first-order bulk term: $\philo \rvert_{0} = -\flo \rvert_{0} $. Using Eq.~\ref{eq:fbulk1}, the condition for the left boundary is:

\begin{equation}\label{eq:phiLeftbound}
    \philo\rvert_0 = -c_1 \ezl \left. \frac{\partial g_0}{\partial x} \right|_{0} - \sum_\lambda \epsilon^{-1} \M^{-1} \left. \frac{\partial g_0}{\partial x}\right|_{0} \ezlp
\end{equation}
where, as in Ref.~\cite{Peraud2016a}, we anticipate the boundary term to scale as the gradient of the previous order solution:

\begin{equation}\label{eq:g1Left}
    g_1\rvert_{x=0} = -c_1 \left. \frac{\partial g_0}{\partial \xbar} \right \rvert_{x=0}
\end{equation}

An analogous boundary condition applies to the right boundary. The unknown constant $c_1$ captures the jump-type boundary condition for non-diffusive corrections at  first order. Solving for $c_1$ involves obtaining the eigenvalues and eigenvectors of $M$ and is described in detail in Ref.~\cite{Peraud2016a}. In brief, the general solution to Eq.~\ref{eq:phi1} can be written as a linear combination of the eigenvectors corresponding to the negative eigenvalues of $M$:

\begin{equation}
    \philo= \sum_{i} A_i h_{\lambda,i} e^{\rho_i \eta} 
\end{equation}

where $A_i$ are unknown coefficients, and $\rho_i$ and $h_{\lambda,i}$ are the negative eigenvalues and the corresponding eigenvectors, respectively. Only the negative eigenvalues are used so that the boundary solution tends to zero in the bulk. Then, Eq.~\ref{eq:phiLeftbound} gives a linear system of equations for $A_i$ and $c_1$. We note that in the formulation in Ref.~\cite{Peraud2016a}, $c_1$ represents a deviation in a linearized quantity, such as temperature, relative to an equilibrium value. However,  in the present formulation, $c_1$ multiplies the absolute distribution function rather than a deviation and is not restricted to linear deviations from that distribution.

After calculating $c_1$, the first-order solution is completely specified using the diffusion equation for $g_1(x)$ and the boundary condition from Eq.~\ref{eq:g1Left}. The analysis progresses similarly for higher-order solutions. For the specific slab problem here, it can be shown that the jump coefficients associated with second-order derivatives vanish as in Appendix C of Ref.~\cite{Peraud2016a}. Therefore, $c_1$ is the only required coefficient, allowing the asymptotic expansion to be summed over all higher-order terms. After summing $\fl$ over the Brillouin zone to obtain the carrier density, we obtain:

\begin{equation}\label{eq:nx}
    n(x) = n_0 + (n_R - n_L)(1-2 \xbar) \sum_{n=1}^\infty \epsilon^n (-2)^{n-1} c_1^n = n_0 + \frac{\epsilon c_1}{1 + 2\epsilon c_1} (n_R - n_L)(1-2\xbar)
\end{equation}
\noindent
where $n_0(x)$ is the zeroth-order carrier density, and the higher-order terms correct for non-diffusive transport. Because the macroscopic constitutive relation between electric current density and number density gradient applies in the bulk, \cite{Peraud2016a} the current density including the effect of quasiballistic transport is given by:

\begin{equation}\label{eq:diffCurrent}
    J_{e,x} = -q D_{e,Si} \frac{\partial n(x)}{\partial x}
\end{equation}

Here, $D_{e,Si}$ is the bulk diffusivity. These results have been extensively validated by comparison to Monte Carlo simulations in Ref.~\cite{Peraud2016a}. We also observe from Eq.~\ref{eq:nx} that the dependence of carrier density and thus electric current on base-emitter voltage $V_{BE}$ is only through the left boundary condition $n_{L} \sim e^{qV_{BE}/kT}$, allowing the current-voltage characteristics for arbitrary $V_{BE}$ to be obtained once $c_1$ is computed.

We briefly discuss the features of the present asymptotic approach compared to those reported previously for electronic transport. An early study of the slab electron transport problem employed a direct scattering matrix solution to the Boltzmann equation \cite{Stettler1994} that required discretization in wave vector and spatial coordinates and was thus numerically expensive. Approximate methods such as the McKelvey flux method \cite{Tanaka, Maassen2015a} and the Landauer approach \cite{Maassen2015, Lundstrom2018} were next applied for electron and phonon transport. These flux-based methods provide approximate solutions to the Boltzmann equation with minimal computational expense due to simplifications to the collision term. The advantage of the present approach is that it provides the exact solution to the Boltzmann equation with the full collision matrix while requiring only discretization in wave vector space rather than in both real space and wave vector space. The resulting calculation is thus orders of magnitude faster than the purely numerical solution of the Boltzmann equation. Further, compared to Ref.~\cite{Peraud2016a} this work does not require linearization around an equilibrium distribution and can accommodate arbitrary crystals, rather than isotropic ones; and the full collision matrix, rather than the relaxation time approximation. Finally, with a standard numerical treatment, the current density would need to be recomputed when any variable is changed. With the present method, the expensive calculation for $c_1$ only needs to be performed once at each temperature because the average Knudsen number and $V_{BE}$ are independent of $c_1$ and included directly in Eq.~\ref{eq:nx}.

\section{Results}

We apply this approach to study 1D steady-state electron transport in slabs of non-degenerate silicon ($N_A = 3 \times 10^{-18}$ cm$^{-3}$) and of width $L \sim 40$ nm - $100$ $\mu$m. The hemispherical distribution function at the left and right boundaries is a Boltzmann distribution with carrier density $n_L = n_{L,eq} \exp{(q V_{BE}/kT)}$ and $n_R=0$, respectively, where $n_{L,eq}$ is the equilibrium minority carrier concentration. Note that at high biases the law of the junction used for $n_L$ may not be strictly satisfied due to degenerate carrier statistics, but this approximation will not affect our conclusions as discussed later. The junction is assumed to be a step junction with a uniform Ge fraction of 0.2, with bandgap values for SiGe obtained from Ref.~\cite{ioffeSi}. Owing to lack of precise knowledge of the electronic structure and scattering mechanisms  in the strained SiGe:C films used in devices,  in this work we assume parabolic bands and employ the relaxation time approximation, with energy-dependent impurity and acoustic phonon scattering rates based on the forms given in Chapter 3 of Ref.~\cite{Lundstrom2000fundamentals}. Several works report that the temperature dependence of minority carrier mobility in p-doped SiGe does not obey the $\sim T^{1.5}$ scaling typical of impurity scattering at cryogenic temperatures \cite{Swirhun1988,Klaassenn1992,Rieh2000a}. Therefore, we extract coefficients and exponents for relaxation time $\tau =  \mu_0 T^{p} E^{s}$ by fitting the computed mobility to minority carrier mobility data reported in Fig.~3 of Ref.~\cite{Rieh2000a} and combining the relaxation times for acoustic and impurity scattering using Matthiessen's rule. The fitting coefficients and exponents for acoustic (impurity) scattering rates are $\mu_0 = 350~ (700)$ cm$ ^{2}$V$ ^{-1}$s$ ^{-1}$, $p=-1.5~ (0.25)$ and $s = -0.5~ (-0.5)$, respectively. Our conclusions are robust to these assumptions. We solve for $c_1$ at the temperatures corresponding to Ref.~\cite{Rucker2017}, assuming an isotropic crystal using a $2200 \times 2200$ grid in energy and angular coordinate space. The collision matrix used in this work  corresponds to Eq.~B.5 of Ref.~\cite{Peraud2016a}. For example, at 300 K we find $c_1 = 0.7341$.

\begin{figure}[h]\label{fig1:nxJKn}
    \centering{\phantomsubcaption \label{fig1:nxJKn_a}
    \phantomsubcaption \label{fig1:nxJKn_b}}
    \includegraphics[width = \textwidth]{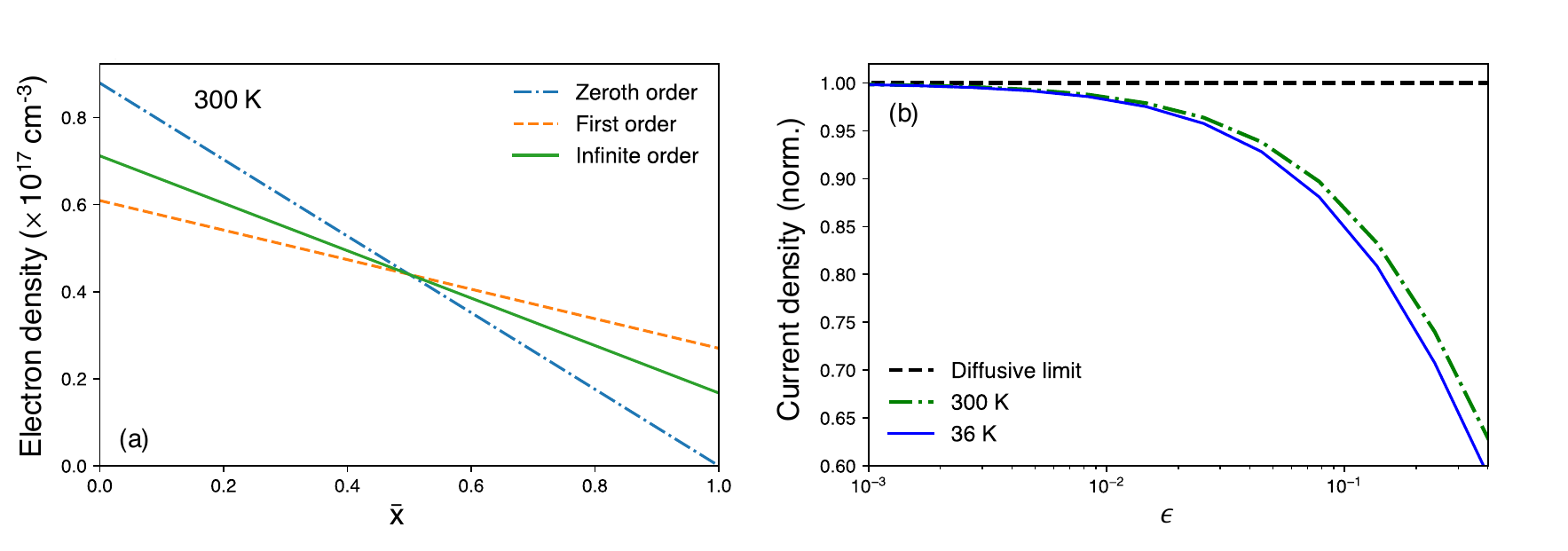}
    \caption{(a) Electron density $n(x)$ versus normalized distance $\xbar$ at 300 K and $V_{BE}=0.8$ V at zeroth order (blue dash-dotted line), first order (orange dashed line) and infinite order (green solid line). (b) Normalized electric current density versus Knudsen number $\epsilon$ at 300 K (green dash-dotted line) and 36 K (blue solid line) derived from carrier concentration gradient. The current density is normalized to the diffusion current density obtained at zeroth order (dashed line). Quasiballistic transport predicts a comparable decrease in current relative to drift-diffusion at both room and cryogenic temperatures as Knudsen number increases.}
        
\end{figure}

Figure \ref{fig1:nxJKn_a} shows the carrier density versus normalized distance at 300 K for  $L = 40$ nm and $V_{BE}=0.8$ V calculated at different orders of the expansion. The zeroth-order solution is by definition the drift-diffusion solution and exhibits a linear profile with a carrier density at the left edge of the slab of $n_L = n_{L,eq} \exp{(q V_{BE}/kT)} = 0.87 \times 10^{17}$ cm$^{-3}$. The higher-order non-diffusive corrections predict a reduced carrier concentration gradient across the base relative to the zeroth-order case. This feature occurs because at $\epsilon= 0.38$, transport is quasiballistic, with electron mean free paths being on the order of the base width. Under these conditions, the ballistic propagation of some electrons across the base lowers the carrier density at the left boundary and increases it at the right boundary, resulting in jump-type boundary conditions at each edge and a shallower carrier density across the base. The solution at the first-order over-corrects for non-diffusive effects, but the infinite series accounts for higher-order terms and lies in between the zeroth and first-order solutions. This non-monotonicity occurs because the closed form solution switches signs at each order in Eq.~\ref{eq:nx}.

To study how the electric current density is impacted by quasiballistic transport, we calculate the current density versus Knudsen number in Fig.~\ref{fig1:nxJKn_b} for two temperatures values reported in Ref.~\cite{Rucker2017}. The electric current density from Eq.~\ref{eq:diffCurrent} is proportional to the carrier concentration gradient in Fig.~\ref{fig1:nxJKn_a} for a given voltage, temperature and Knudsen number. At 300 K, the current density is observed to decrease from the diffusive limit as $\epsilon \rightarrow 1$ (quasiballistic limit). As seen from Fig.~\ref{fig1:nxJKn_a}, quasiballistic corrections lower the concentration gradient, and the contribution of the correction terms is multiplied by Knudsen number, resulting in greater deviation with increasing Knudsen number. Therefore, a shallower concentration gradient and consequently a decreased current density is expected with increasing Knudsen number. At 36 K corresponding to the temperature reported in Ref.~\cite{Rucker2017}, the calculation for $c_1$ is repeated to give a value of $0.733$. At this temperature, we see that the deviation from the diffusive limit is comparable to that at 300 K. The difference between the two curves is due to changes in bulk diffusivity at 36 K that are independent of $c_1$ and predict a slightly smaller current density than at 300 K. This calculation shows that even at cryogenic temperatures, quasiballistic effects yield a lower collector current density relative to the drift-diffusion prediction, opposite to that observed experimentally.

\begin{figure}[ht]\label{fig2:JVgmT}
    \centering{\phantomsubcaption \label{fig2:JVgmT_a}
        \phantomsubcaption \label{fig2:JVgmT_b}}
    \includegraphics[width = \textwidth]{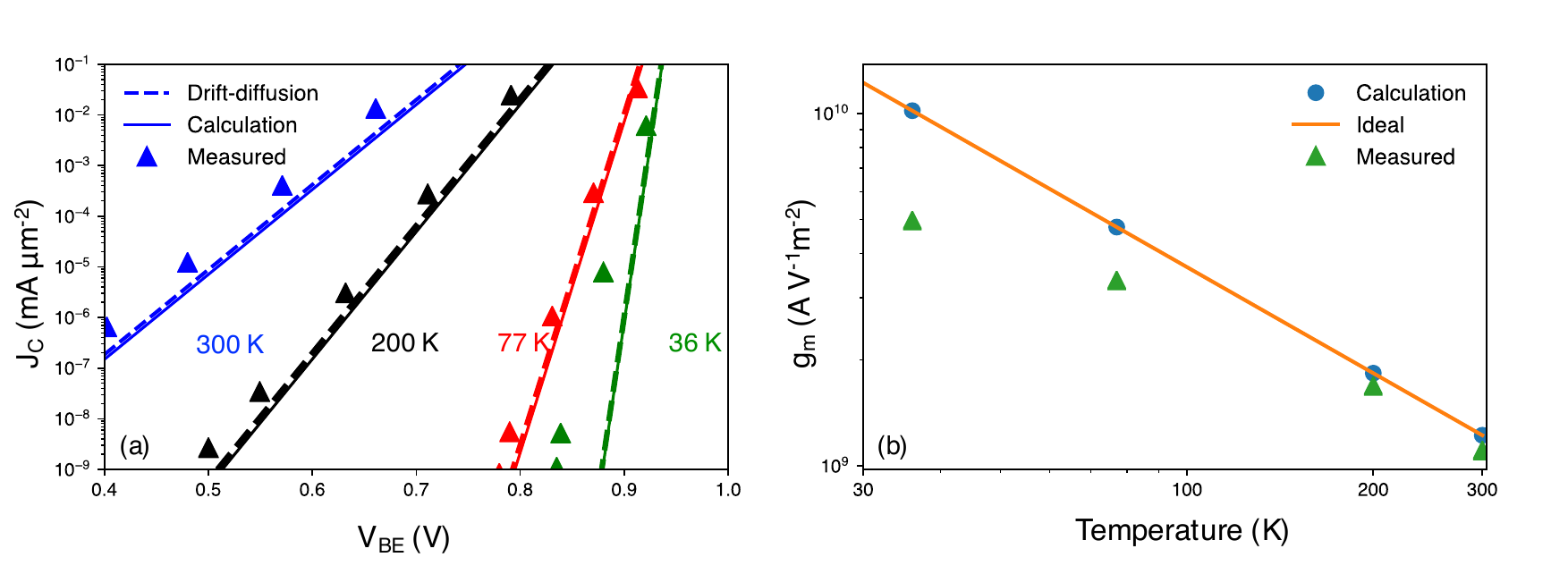}
    \caption{(a) Collector current density $J_c$ versus base-emitter voltage $V_{BE}$ from drift-diffusion (dashed line) and quasiballistic calculations (solid line) compared to measured data (triangles, Ref.~\cite{Rucker2017}) at various temperatures. (b) Transconductance per unit area $g_m$ versus temperature for an ideal diode (dashed line), from calculation (circles), and experiment (triangles, Ref.~\cite{Rucker2017}) for a fixed $J_C$ derived from the $J-V$ characteristics in (a). The present calculations do not predict the increased current density or decreased transconductance observed in measurements at cryogenic temperatures.}
\end{figure}

We now compare the predictions of our calculation with experimentally reported DC IV characteristics. The data from Ref.~\cite{Rucker2017} is selected due to the availability of Gummel curves for a state-of-the-art device at a reasonably dense concentration of temperatures. Figure~\ref{fig2:JVgmT_a} plots calculated and measured collector current density versus base-emitter voltage $V_{BE}$. We observe that at all temperatures, quasiballistic transport predicts a weakly reduced current density compared to the  drift-diffusion prediction, as expected from the discussion of Fig.~\ref{fig1:nxJKn_b}. The calculation nearly coincide in Fig.~\ref{fig2:JVgmT_a} because the deviation of quasiballistic current density from the diffusive current density is small on the logarithmic scale in Fig.~\ref{fig2:JVgmT_a}. At 300 K, 200 K, and 77 K, both predictions agree well with measured data. However, at 36 K, quasiballistic and drift-diffusion current density predictions are similar in magnitude but are both orders of magnitude lower than the measured data for a given $V_{BE}$. Further, the predicted slope of the $J-V$ curve is unchanged from the drift-diffusion prediction and does not saturate to a value similar to that at 77 K as observed in the measured data.

To further examine this result, we present transconductance versus temperature at a fixed value of collector current density in Fig.~\ref{fig2:JVgmT_b}. The transconductance per unit area $g_m$ is calculated as the slope of the current-voltage curve at a fixed bias, and for an ideal SiGe HBT it is given by $g_m = q J_C/k T$. The measured transconductance exhibits a plateau to a lower value than the ideal value as temperature decreases. However, the transconductance values calculated from the slope of the calculated results in Fig.~\ref{fig2:JVgmT_a} are identical to the ideal values. Therefore, quasiballistic transport does not change the temperature dependence of the transconductance. The reason can be seen in Eq.~\ref{eq:nx} in which  $V_{BE}$ affects the electron density only through the left  boundary condition $n_L$; therefore, a change in voltage affects current density in a way that is independent of the mechanism of carrier transport and thus preserves the diffusive temperature dependence regardless of the Knudsen number.

\section{Discussion}

Our analysis has showed that the predicted influence of quasiballistic transport on the IV characteristics of HBTs is inconsistent with the experimental observations. We now address several mechanisms that we have not included in our analysis. First, we have not incorporated the built-in field in the base region due to Ge grading. However, as in Ref.~\cite{choi2021electronic}, the only effect of a field term is to modify $M$ and thus  change the value of $c_1$, and our conclusions are robust to such changes. Second, modern devices may have base doping levels exceeding $10^{19}$ $cm^{-3}$ \cite{Schroter2011,Rucker2017,Ying2018} to minimize base resistance and prevent carrier freeze-out. This high level of doping leads to degenerate carrier statistics, while we have assumed non-degenerate statistics. However, non-idealities in the IV characteristics are observed at temperatures up to $\sim 80$ K for which the electrons are non-degenerate, and the present analysis thus applies under relevant conditions. Further, in degenerate conditions the law of the junction overpredicts the minority carrier density,~\cite{jainRodwell2011} implying that accounting for degenerate statistics would predict a further decreased collector current that again contradicts experiment. Lastly, the asymptotic solution is restricted to values of average Knudsen number such that  $2\epsilon c_1 < 1$, which limits our analysis to a minimum base width of 40 nm for the chosen scattering rates. However, DC non-idealities were reported in devices with base widths on this order,\cite{Davidovic2017, Rucker2017, Ying2018} and again our analysis applies. Given these considerations, we conclude that quasiballistic transport is not responsible for non-ideal cryogenic DC characteristics of SiGe HBTs.

Finally, we discuss alternate explanations for the cryogenic non-idealities. Prior works have suggested that  direct tunneling from the emitter to the collector is a possible origin of non-idealities in devices with narrow base widths $\sim 10$ $nm$ \cite{Rucker2017,Davidovic2017}. However, non-ideal cryogenic transconductance has been observed in first-generation SiGe HBTs with base widths on the order of $100$ nm \cite{Richey1996, Ying2018} for which the direct tunneling current is expected to be negligible. Other reported evidence for a transition from trap-assisted transport to a tunneling or quasiballistic mechanism is the change in slope of the collector I-V characteristics \cite{Ying2018,Davidovic2017}. However, similar trends are observed in the forward regime of Schottky diodes \cite{Tung1992,ewing2007} for which direct tunneling or quasiballistic transport is not relevant, and the possibility of a similar mechanism occurring in SiGe HBTs has not been excluded.

We instead offer a hypothesis for the origin of cryogenic non-idealities as originating from spatial inhomogeneities in the base-emitter potential barrier. The earliest works on heavily doped p-n junctions reported temperature-independent slopes of cryogenic forward-bias I-V characteristics for voltages where band-to-band tunneling is unlikely to occur \cite{esaki1960, Dumin1965,delAlamo1986}. Extensive studies of Schottky diodes have reported a variety of cryogenic $I-V$ non-idealities including the so-called $T_0$ anomaly \cite{padovani1965experimental} and temperature-dependent ideality factors \cite{padovani1966field,hackam1972electrical,Bhuiyan1988}. These non-idealities have been explained by inhomogeneities in barrier height \cite{Tung1992} that exist even in high-quality epitaxial junctions such as NiSi$_2$/Si junctions and have been linked to the atomic structure of the interface \cite{Tung1984Nisi,Tung1991nisi}. In SiGe HBT junctions, non-ideal base currents and the resulting random telegraph noise have been attributed to voltage barrier height fluctuations arising from trap states in the base-emitter space charge region \cite{vonHaartman2002rts} such as those of electrically-active carbon defects \cite{raoult2008}. Future work will examine whether such barrier inhomogeneities are capable of explaining the collector current non-idealities of SiGe HBTs.

\section{Conclusion}
We have reported a study of quasiballistic transport in SiGe HBTs using an exact, semi-analytic solution to the Boltzmann equation based on an asymptotic expansion method. We find that the predicted IV characteristics including quasiballistic transport are inconsistent with experiment. Specifically, our calculations including quasiballistic transport predict collector currents that are orders of magnitude smaller than the measured currents for a given base-emitter voltage and an unaltered temperature dependence of transconductance relative to the ideal value, both of which contradict experimental observations. We suggest that local fluctuations in the base-emitter barrier height could account for the non-ideal collector current as has been observed in Schottky diodes. Our work advances the understanding of electronic transport in cryogenic SiGe HBTs that is required to further improve their microwave noise performance.

\begin{acknowledgements}
The authors thank Mark Lundstrom, J.P.~Peraud, and Nicolas Hadjiconstantinou for useful discussions. This work was supported by NSF Award Number 1911926. 
\end{acknowledgements}

The data that support the findings of this study are available from the corresponding author upon reasonable request.

\bibliography{AsymptoticBE.bib}

\begin{thebibliography}{66}%
\makeatletter
\providecommand \@ifxundefined [1]{%
 \@ifx{#1\undefined}
}%
\providecommand \@ifnum [1]{%
 \ifnum #1\expandafter \@firstoftwo
 \else \expandafter \@secondoftwo
 \fi
}%
\providecommand \@ifx [1]{%
 \ifx #1\expandafter \@firstoftwo
 \else \expandafter \@secondoftwo
 \fi
}%
\providecommand \natexlab [1]{#1}%
\providecommand \enquote  [1]{``#1''}%
\providecommand \bibnamefont  [1]{#1}%
\providecommand \bibfnamefont [1]{#1}%
\providecommand \citenamefont [1]{#1}%
\providecommand \href@noop [0]{\@secondoftwo}%
\providecommand \href [0]{\begingroup \@sanitize@url \@href}%
\providecommand \@href[1]{\@@startlink{#1}\@@href}%
\providecommand \@@href[1]{\endgroup#1\@@endlink}%
\providecommand \@sanitize@url [0]{\catcode `\\12\catcode `\$12\catcode
  `\&12\catcode `\#12\catcode `\^12\catcode `\_12\catcode `\%12\relax}%
\providecommand \@@startlink[1]{}%
\providecommand \@@endlink[0]{}%
\providecommand \url  [0]{\begingroup\@sanitize@url \@url }%
\providecommand \@url [1]{\endgroup\@href {#1}{\urlprefix }}%
\providecommand \urlprefix  [0]{URL }%
\providecommand \Eprint [0]{\href }%
\providecommand \doibase [0]{https://doi.org/}%
\providecommand \selectlanguage [0]{\@gobble}%
\providecommand \bibinfo  [0]{\@secondoftwo}%
\providecommand \bibfield  [0]{\@secondoftwo}%
\providecommand \translation [1]{[#1]}%
\providecommand \BibitemOpen [0]{}%
\providecommand \bibitemStop [0]{}%
\providecommand \bibitemNoStop [0]{.\EOS\space}%
\providecommand \EOS [0]{\spacefactor3000\relax}%
\providecommand \BibitemShut  [1]{\csname bibitem#1\endcsname}%
\let\auto@bib@innerbib\@empty
%</preamble>
\bibitem [{\citenamefont {Rücker}\ \emph {et~al.}(2012)\citenamefont
  {Rücker}, \citenamefont {Heinemann},\ and\ \citenamefont
  {Fox}}]{rucker2012halfthz}%
  \BibitemOpen
  \bibfield  {author} {\bibinfo {author} {\bibfnamefont {H.}~\bibnamefont
  {Rücker}}, \bibinfo {author} {\bibfnamefont {B.}~\bibnamefont {Heinemann}},\
  and\ \bibinfo {author} {\bibfnamefont {A.}~\bibnamefont {Fox}},\ }\bibfield
  {title} {\bibinfo {title} {{Half-Terahertz SiGe BiCMOS technology}},\ }in\
  \href {https://doi.org/10.1109/SiRF.2012.6160164} {\emph {\bibinfo
  {booktitle} {2012 IEEE 12th Topical Meeting on Silicon Monolithic Integrated
  Circuits in RF Systems}}}\ (\bibinfo {year} {2012})\ pp.\ \bibinfo {pages}
  {133--136}\BibitemShut {NoStop}%
\bibitem [{\citenamefont {Chevalier}\ \emph {et~al.}(2017)\citenamefont
  {Chevalier}, \citenamefont {Schroter}, \citenamefont {Bolognesi},
  \citenamefont {D'Alessandro}, \citenamefont {Alexandrova}, \citenamefont
  {Bock}, \citenamefont {Flickiger}, \citenamefont {Fregonese}, \citenamefont
  {Heinemann}, \citenamefont {Jungemann}, \citenamefont {Lovblom},
  \citenamefont {Maneux}, \citenamefont {Ostinelli}, \citenamefont {Pawlak},
  \citenamefont {Rinaldi}, \citenamefont {Rucker}, \citenamefont {Wedel},\ and\
  \citenamefont {Zimmer}}]{Chevalier2017}%
  \BibitemOpen
  \bibfield  {author} {\bibinfo {author} {\bibfnamefont {P.}~\bibnamefont
  {Chevalier}}, \bibinfo {author} {\bibfnamefont {M.}~\bibnamefont {Schroter}},
  \bibinfo {author} {\bibfnamefont {C.~R.}\ \bibnamefont {Bolognesi}}, \bibinfo
  {author} {\bibfnamefont {V.}~\bibnamefont {D'Alessandro}}, \bibinfo {author}
  {\bibfnamefont {M.}~\bibnamefont {Alexandrova}}, \bibinfo {author}
  {\bibfnamefont {J.}~\bibnamefont {Bock}}, \bibinfo {author} {\bibfnamefont
  {R.}~\bibnamefont {Flickiger}}, \bibinfo {author} {\bibfnamefont
  {S.}~\bibnamefont {Fregonese}}, \bibinfo {author} {\bibfnamefont
  {B.}~\bibnamefont {Heinemann}}, \bibinfo {author} {\bibfnamefont
  {C.}~\bibnamefont {Jungemann}}, \bibinfo {author} {\bibfnamefont
  {R.}~\bibnamefont {Lovblom}}, \bibinfo {author} {\bibfnamefont
  {C.}~\bibnamefont {Maneux}}, \bibinfo {author} {\bibfnamefont
  {O.}~\bibnamefont {Ostinelli}}, \bibinfo {author} {\bibfnamefont
  {A.}~\bibnamefont {Pawlak}}, \bibinfo {author} {\bibfnamefont
  {N.}~\bibnamefont {Rinaldi}}, \bibinfo {author} {\bibfnamefont
  {H.}~\bibnamefont {Rucker}}, \bibinfo {author} {\bibfnamefont
  {G.}~\bibnamefont {Wedel}},\ and\ \bibinfo {author} {\bibfnamefont
  {T.}~\bibnamefont {Zimmer}},\ }\bibfield  {title} {\bibinfo {title}
  {{Si/SiGe:C and InP/GaAsSb Heterojunction Bipolar Transistors for THz
  Applications}},\ }\href {https://doi.org/10.1109/JPROC.2017.2669087}
  {\bibfield  {journal} {\bibinfo  {journal} {Proceedings of the IEEE}\
  }\textbf {\bibinfo {volume} {105}},\ \bibinfo {pages} {1035} (\bibinfo {year}
  {2017})}\BibitemShut {NoStop}%
\bibitem [{\citenamefont {Bardin}\ \emph {et~al.}(2013)\citenamefont {Bardin},
  \citenamefont {Ravindran}, \citenamefont {Chang}, \citenamefont {Kumar},
  \citenamefont {Stern}, \citenamefont {Shaw}, \citenamefont {Russell},\ and\
  \citenamefont {Farr}}]{bardin2013high}%
  \BibitemOpen
  \bibfield  {author} {\bibinfo {author} {\bibfnamefont {J.~C.}\ \bibnamefont
  {Bardin}}, \bibinfo {author} {\bibfnamefont {P.}~\bibnamefont {Ravindran}},
  \bibinfo {author} {\bibfnamefont {S.-W.}\ \bibnamefont {Chang}}, \bibinfo
  {author} {\bibfnamefont {R.}~\bibnamefont {Kumar}}, \bibinfo {author}
  {\bibfnamefont {J.~A.}\ \bibnamefont {Stern}}, \bibinfo {author}
  {\bibfnamefont {M.~D.}\ \bibnamefont {Shaw}}, \bibinfo {author}
  {\bibfnamefont {D.~S.}\ \bibnamefont {Russell}},\ and\ \bibinfo {author}
  {\bibfnamefont {W.~H.}\ \bibnamefont {Farr}},\ }\bibfield  {title} {\bibinfo
  {title} {{A high-speed cryogenic SiGe channel combiner IC for large
  photon-starved SNSPD arrays}},\ }in\ \href
  {https://doi.org/10.1109/BCTM.2013.6798179} {\emph {\bibinfo {booktitle}
  {2013 IEEE Bipolar/BiCMOS Circuits and Technology Meeting (BCTM)}}}\
  (\bibinfo {organization} {IEEE},\ \bibinfo {year} {2013})\ pp.\ \bibinfo
  {pages} {215--218}\BibitemShut {NoStop}%
\bibitem [{\citenamefont {Yuan}\ \emph {et~al.}(2009)\citenamefont {Yuan},
  \citenamefont {Cressler}, \citenamefont {Krithivasan}, \citenamefont
  {Thrivikraman}, \citenamefont {Khater}, \citenamefont {Ahlgren},
  \citenamefont {Joseph},\ and\ \citenamefont {Rieh}}]{Yuan2009}%
  \BibitemOpen
  \bibfield  {author} {\bibinfo {author} {\bibfnamefont {J.}~\bibnamefont
  {Yuan}}, \bibinfo {author} {\bibfnamefont {J.~D.}\ \bibnamefont {Cressler}},
  \bibinfo {author} {\bibfnamefont {R.}~\bibnamefont {Krithivasan}}, \bibinfo
  {author} {\bibfnamefont {T.}~\bibnamefont {Thrivikraman}}, \bibinfo {author}
  {\bibfnamefont {M.~H.}\ \bibnamefont {Khater}}, \bibinfo {author}
  {\bibfnamefont {D.~C.}\ \bibnamefont {Ahlgren}}, \bibinfo {author}
  {\bibfnamefont {A.~J.}\ \bibnamefont {Joseph}},\ and\ \bibinfo {author}
  {\bibfnamefont {J.~S.}\ \bibnamefont {Rieh}},\ }\bibfield  {title} {\bibinfo
  {title} {{On the performance limits of cryogenically operated SiGe HBTs and
  its relation to scaling for terahertz speeds}},\ }\href
  {https://doi.org/10.1109/TED.2009.2016017} {\bibfield  {journal} {\bibinfo
  {journal} {IEEE Transactions on Electron Devices}\ }\textbf {\bibinfo
  {volume} {56}},\ \bibinfo {pages} {1007} (\bibinfo {year}
  {2009})}\BibitemShut {NoStop}%
\bibitem [{\citenamefont {Cressler}(2021)}]{cressler2021new}%
  \BibitemOpen
  \bibfield  {author} {\bibinfo {author} {\bibfnamefont {J.~D.}\ \bibnamefont
  {Cressler}},\ }\bibfield  {title} {\bibinfo {title} {{New Developments in
  SiGe HBT Reliability for RF Through mmW Circuits}},\ }in\ \href
  {https://doi.org/10.1109/IRPS46558.2021.9405171} {\emph {\bibinfo {booktitle}
  {2021 IEEE International Reliability Physics Symposium (IRPS)}}}\ (\bibinfo
  {organization} {IEEE},\ \bibinfo {year} {2021})\ pp.\ \bibinfo {pages}
  {1--6}\BibitemShut {NoStop}%
\bibitem [{\citenamefont {Frounchi}\ and\ \citenamefont
  {Cressler}(2020)}]{frounchi2020sige}%
  \BibitemOpen
  \bibfield  {author} {\bibinfo {author} {\bibfnamefont {M.}~\bibnamefont
  {Frounchi}}\ and\ \bibinfo {author} {\bibfnamefont {J.~D.}\ \bibnamefont
  {Cressler}},\ }\bibfield  {title} {\bibinfo {title} {A sige millimeter-wave
  front-end for remote sensing and imaging},\ }in\ \href
  {https://doi.org/10.1109/RFIC49505.2020.9218399} {\emph {\bibinfo {booktitle}
  {2020 IEEE Radio Frequency Integrated Circuits Symposium (RFIC)}}}\ (\bibinfo
  {year} {2020})\ pp.\ \bibinfo {pages} {227--230}\BibitemShut {NoStop}%
\bibitem [{\citenamefont {Cressler}(2005)}]{cressler2005potential}%
  \BibitemOpen
  \bibfield  {author} {\bibinfo {author} {\bibfnamefont {J.~D.}\ \bibnamefont
  {Cressler}},\ }\bibfield  {title} {\bibinfo {title} {{On the potential of
  SiGe HBTs for extreme environment electronics}},\ }\href
  {https://doi.org/10.1109/JPROC.2005.852225} {\bibfield  {journal} {\bibinfo
  {journal} {Proceedings of the IEEE}\ }\textbf {\bibinfo {volume} {93}},\
  \bibinfo {pages} {1559} (\bibinfo {year} {2005})}\BibitemShut {NoStop}%
\bibitem [{\citenamefont {Cressler}\ and\ \citenamefont
  {Niu}(2003)}]{cressler2003silicon}%
  \BibitemOpen
  \bibfield  {author} {\bibinfo {author} {\bibfnamefont {J.~D.}\ \bibnamefont
  {Cressler}}\ and\ \bibinfo {author} {\bibfnamefont {G.}~\bibnamefont {Niu}},\
  }\href@noop {} {\emph {\bibinfo {title} {{Silicon-germanium heterojunction
  bipolar transistors}}}}\ (\bibinfo  {publisher} {Artech house},\ \bibinfo
  {year} {2003})\BibitemShut {NoStop}%
\bibitem [{\citenamefont {Heinemann}\ \emph {et~al.}(2017)\citenamefont
  {Heinemann}, \citenamefont {Rucker}, \citenamefont {Barth}, \citenamefont
  {Barwolf}, \citenamefont {Drews}, \citenamefont {Fischer}, \citenamefont
  {Fox}, \citenamefont {Fursenko}, \citenamefont {Grabolla}, \citenamefont
  {Herzel}, \citenamefont {Katzer}, \citenamefont {Korn}, \citenamefont
  {Kruger}, \citenamefont {Kulse}, \citenamefont {Lenke}, \citenamefont
  {Lisker}, \citenamefont {Marschmeyer}, \citenamefont {Scheit}, \citenamefont
  {Schmidt}, \citenamefont {Schmidt}, \citenamefont {Schubert}, \citenamefont
  {Trusch}, \citenamefont {Wipf},\ and\ \citenamefont
  {Wolansky}}]{Heinemann2017}%
  \BibitemOpen
  \bibfield  {author} {\bibinfo {author} {\bibfnamefont {B.}~\bibnamefont
  {Heinemann}}, \bibinfo {author} {\bibfnamefont {H.}~\bibnamefont {Rucker}},
  \bibinfo {author} {\bibfnamefont {R.}~\bibnamefont {Barth}}, \bibinfo
  {author} {\bibfnamefont {F.}~\bibnamefont {Barwolf}}, \bibinfo {author}
  {\bibfnamefont {J.}~\bibnamefont {Drews}}, \bibinfo {author} {\bibfnamefont
  {G.~G.}\ \bibnamefont {Fischer}}, \bibinfo {author} {\bibfnamefont
  {A.}~\bibnamefont {Fox}}, \bibinfo {author} {\bibfnamefont {O.}~\bibnamefont
  {Fursenko}}, \bibinfo {author} {\bibfnamefont {T.}~\bibnamefont {Grabolla}},
  \bibinfo {author} {\bibfnamefont {F.}~\bibnamefont {Herzel}}, \bibinfo
  {author} {\bibfnamefont {J.}~\bibnamefont {Katzer}}, \bibinfo {author}
  {\bibfnamefont {J.}~\bibnamefont {Korn}}, \bibinfo {author} {\bibfnamefont
  {A.}~\bibnamefont {Kruger}}, \bibinfo {author} {\bibfnamefont
  {P.}~\bibnamefont {Kulse}}, \bibinfo {author} {\bibfnamefont
  {T.}~\bibnamefont {Lenke}}, \bibinfo {author} {\bibfnamefont
  {M.}~\bibnamefont {Lisker}}, \bibinfo {author} {\bibfnamefont
  {S.}~\bibnamefont {Marschmeyer}}, \bibinfo {author} {\bibfnamefont
  {A.}~\bibnamefont {Scheit}}, \bibinfo {author} {\bibfnamefont
  {D.}~\bibnamefont {Schmidt}}, \bibinfo {author} {\bibfnamefont
  {J.}~\bibnamefont {Schmidt}}, \bibinfo {author} {\bibfnamefont {M.~A.}\
  \bibnamefont {Schubert}}, \bibinfo {author} {\bibfnamefont {A.}~\bibnamefont
  {Trusch}}, \bibinfo {author} {\bibfnamefont {C.}~\bibnamefont {Wipf}},\ and\
  \bibinfo {author} {\bibfnamefont {D.}~\bibnamefont {Wolansky}},\ }\bibfield
  {title} {\bibinfo {title} {{SiGe HBT with fx/fmax of 505 GHz/720 GHz}},\
  }\href {https://doi.org/10.1109/IEDM.2016.7838335} {\bibfield  {journal}
  {\bibinfo  {journal} {Technical Digest - International Electron Devices
  Meeting, IEDM}\ }\textbf {\bibinfo {volume} {2}},\ \bibinfo {pages} {3.1.1}
  (\bibinfo {year} {2017})}\BibitemShut {NoStop}%
\bibitem [{\citenamefont {Bardin}\ \emph {et~al.}(2017)\citenamefont {Bardin},
  \citenamefont {Montazeri},\ and\ \citenamefont {Chang}}]{Bardin2017}%
  \BibitemOpen
  \bibfield  {author} {\bibinfo {author} {\bibfnamefont {J.~C.}\ \bibnamefont
  {Bardin}}, \bibinfo {author} {\bibfnamefont {S.}~\bibnamefont {Montazeri}},\
  and\ \bibinfo {author} {\bibfnamefont {S.~W.}\ \bibnamefont {Chang}},\
  }\bibfield  {title} {\bibinfo {title} {{Silicon Germanium Cryogenic Low Noise
  Amplifiers}},\ }\bibfield  {journal} {\bibinfo  {journal} {Journal of
  Physics: Conference Series}\ }\textbf {\bibinfo {volume} {834}},\ \href
  {https://doi.org/10.1088/1742-6596/834/1/012007}
  {10.1088/1742-6596/834/1/012007} (\bibinfo {year} {2017})\BibitemShut
  {NoStop}%
\bibitem [{\citenamefont {Bardin}(2009)}]{Bardin2009}%
  \BibitemOpen
  \bibfield  {author} {\bibinfo {author} {\bibfnamefont {J.~C.}\ \bibnamefont
  {Bardin}},\ }\emph {\bibinfo {title} {{Silicon-Germanium Heterojunction
  Bipolar Transistors For Extremely Low-Noise Applications}}},\ \href@noop {}
  {Ph.D. thesis},\ \bibinfo  {school} {California Institute of Technology}
  (\bibinfo {year} {2009})\BibitemShut {NoStop}%
\bibitem [{\citenamefont {Van~Haaren}\ \emph {et~al.}(1998)\citenamefont
  {Van~Haaren}, \citenamefont {Regis}, \citenamefont {Llopis}, \citenamefont
  {Escotte}, \citenamefont {Gruhle}, \citenamefont {Mahner}, \citenamefont
  {Plana},\ and\ \citenamefont {Graffeuil}}]{van1998low}%
  \BibitemOpen
  \bibfield  {author} {\bibinfo {author} {\bibfnamefont {B.}~\bibnamefont
  {Van~Haaren}}, \bibinfo {author} {\bibfnamefont {M.}~\bibnamefont {Regis}},
  \bibinfo {author} {\bibfnamefont {O.}~\bibnamefont {Llopis}}, \bibinfo
  {author} {\bibfnamefont {L.}~\bibnamefont {Escotte}}, \bibinfo {author}
  {\bibfnamefont {A.}~\bibnamefont {Gruhle}}, \bibinfo {author} {\bibfnamefont
  {C.}~\bibnamefont {Mahner}}, \bibinfo {author} {\bibfnamefont
  {R.}~\bibnamefont {Plana}},\ and\ \bibinfo {author} {\bibfnamefont
  {J.}~\bibnamefont {Graffeuil}},\ }\bibfield  {title} {\bibinfo {title}
  {{Low-frequency noise properties of SiGe HBT's and application to ultra-low
  phase-noise oscillators}},\ }\href {https://doi.org/10.1109/22.668677}
  {\bibfield  {journal} {\bibinfo  {journal} {IEEE Transactions on Microwave
  Theory and Techniques}\ }\textbf {\bibinfo {volume} {46}},\ \bibinfo {pages}
  {647} (\bibinfo {year} {1998})}\BibitemShut {NoStop}%
\bibitem [{\citenamefont {Montazeri}\ \emph {et~al.}(2016)\citenamefont
  {Montazeri}, \citenamefont {Wong}, \citenamefont {Coskun},\ and\
  \citenamefont {Bardin}}]{Montazeri2016}%
  \BibitemOpen
  \bibfield  {author} {\bibinfo {author} {\bibfnamefont {S.}~\bibnamefont
  {Montazeri}}, \bibinfo {author} {\bibfnamefont {W.~T.}\ \bibnamefont {Wong}},
  \bibinfo {author} {\bibfnamefont {A.~H.}\ \bibnamefont {Coskun}},\ and\
  \bibinfo {author} {\bibfnamefont {J.~C.}\ \bibnamefont {Bardin}},\ }\bibfield
   {title} {\bibinfo {title} {{Ultra-Low-Power Cryogenic SiGe Low-Noise
  Amplifiers: Theory and Demonstration}},\ }\href
  {https://doi.org/10.1109/TMTT.2015.2497685} {\bibfield  {journal} {\bibinfo
  {journal} {IEEE Transactions on Microwave Theory and Techniques}\ }\textbf
  {\bibinfo {volume} {64}},\ \bibinfo {pages} {178} (\bibinfo {year}
  {2016})}\BibitemShut {NoStop}%
\bibitem [{\citenamefont {Wong}\ \emph {et~al.}(2020)\citenamefont {Wong},
  \citenamefont {Hosseini}, \citenamefont {Rucker},\ and\ \citenamefont
  {Bardin}}]{Wong2020}%
  \BibitemOpen
  \bibfield  {author} {\bibinfo {author} {\bibfnamefont {W.~T.}\ \bibnamefont
  {Wong}}, \bibinfo {author} {\bibfnamefont {M.}~\bibnamefont {Hosseini}},
  \bibinfo {author} {\bibfnamefont {H.}~\bibnamefont {Rucker}},\ and\ \bibinfo
  {author} {\bibfnamefont {J.~C.}\ \bibnamefont {Bardin}},\ }\bibfield  {title}
  {\bibinfo {title} {{A 1 mW cryogenic LNA exploiting optimized SiGe HBTs to
  achieve an average noise temperature of 3.2 K from 4-8 GHz}},\ }\href
  {https://doi.org/10.1109/IMS30576.2020.9224049} {\bibfield  {journal}
  {\bibinfo  {journal} {IEEE MTT-S International Microwave Symposium Digest}\
  }\textbf {\bibinfo {volume} {2020-August}},\ \bibinfo {pages} {181} (\bibinfo
  {year} {2020})}\BibitemShut {NoStop}%
\bibitem [{\citenamefont {Patra}\ \emph {et~al.}(2018)\citenamefont {Patra},
  \citenamefont {Incandela}, \citenamefont {{Van Dijk}}, \citenamefont
  {Homulle}, \citenamefont {Song}, \citenamefont {Shahmohammadi}, \citenamefont
  {Staszewski}, \citenamefont {Vladimirescu}, \citenamefont {Babaie},
  \citenamefont {Sebastiano},\ and\ \citenamefont {Charbon}}]{Patra2018}%
  \BibitemOpen
  \bibfield  {author} {\bibinfo {author} {\bibfnamefont {B.}~\bibnamefont
  {Patra}}, \bibinfo {author} {\bibfnamefont {R.~M.}\ \bibnamefont
  {Incandela}}, \bibinfo {author} {\bibfnamefont {J.~P.}\ \bibnamefont {{Van
  Dijk}}}, \bibinfo {author} {\bibfnamefont {H.~A.}\ \bibnamefont {Homulle}},
  \bibinfo {author} {\bibfnamefont {L.}~\bibnamefont {Song}}, \bibinfo {author}
  {\bibfnamefont {M.}~\bibnamefont {Shahmohammadi}}, \bibinfo {author}
  {\bibfnamefont {R.~B.}\ \bibnamefont {Staszewski}}, \bibinfo {author}
  {\bibfnamefont {A.}~\bibnamefont {Vladimirescu}}, \bibinfo {author}
  {\bibfnamefont {M.}~\bibnamefont {Babaie}}, \bibinfo {author} {\bibfnamefont
  {F.}~\bibnamefont {Sebastiano}},\ and\ \bibinfo {author} {\bibfnamefont
  {E.}~\bibnamefont {Charbon}},\ }\bibfield  {title} {\bibinfo {title}
  {{Cryo-CMOS Circuits and Systems for Quantum Computing Applications}},\
  }\href {https://doi.org/10.1109/JSSC.2017.2737549} {\bibfield  {journal}
  {\bibinfo  {journal} {IEEE Journal of Solid-State Circuits}\ }\textbf
  {\bibinfo {volume} {53}},\ \bibinfo {pages} {309} (\bibinfo {year}
  {2018})}\BibitemShut {NoStop}%
\bibitem [{\citenamefont {Weinreb}\ \emph {et~al.}(2007)\citenamefont
  {Weinreb}, \citenamefont {Bardin},\ and\ \citenamefont {Mani}}]{Weinreb2007}%
  \BibitemOpen
  \bibfield  {author} {\bibinfo {author} {\bibfnamefont {S.}~\bibnamefont
  {Weinreb}}, \bibinfo {author} {\bibfnamefont {J.~C.}\ \bibnamefont
  {Bardin}},\ and\ \bibinfo {author} {\bibfnamefont {H.}~\bibnamefont {Mani}},\
  }\bibfield  {title} {\bibinfo {title} {{Design of cryogenic SiGe low-noise
  amplifiers}},\ }\href {https://doi.org/10.1109/TMTT.2007.907729} {\bibfield
  {journal} {\bibinfo  {journal} {IEEE Transactions on Microwave Theory and
  Techniques}\ }\textbf {\bibinfo {volume} {55}},\ \bibinfo {pages} {2306}
  (\bibinfo {year} {2007})}\BibitemShut {NoStop}%
\bibitem [{\citenamefont {Plana}\ and\ \citenamefont
  {Escotte}(1997)}]{plana1997}%
  \BibitemOpen
  \bibfield  {author} {\bibinfo {author} {\bibfnamefont {R.}~\bibnamefont
  {Plana}}\ and\ \bibinfo {author} {\bibfnamefont {L.}~\bibnamefont
  {Escotte}},\ }\bibfield  {title} {\bibinfo {title} {{Noise properties of
  micro-wave heterojunction bipolar transistors}},\ }in\ \href
  {https://doi.org/10.1109/ICMEL.1997.625221} {\emph {\bibinfo {booktitle}
  {1997 21st International Conference on Microelectronics. Proceedings}}},\
  Vol.~\bibinfo {volume} {1}\ (\bibinfo {year} {1997})\ pp.\ \bibinfo {pages}
  {215--222 vol.1}\BibitemShut {NoStop}%
\bibitem [{\citenamefont {Niu}(2005)}]{Niu2005}%
  \BibitemOpen
  \bibfield  {author} {\bibinfo {author} {\bibfnamefont {G.}~\bibnamefont
  {Niu}},\ }\bibfield  {title} {\bibinfo {title} {{Noise in SiGe HBT RF
  technology: Physics, modeling, and circuit implications}},\ }\href
  {https://doi.org/10.1109/JPROC.2005.852226} {\bibfield  {journal} {\bibinfo
  {journal} {Proceedings of the IEEE}\ }\textbf {\bibinfo {volume} {93}},\
  \bibinfo {pages} {1583} (\bibinfo {year} {2005})}\BibitemShut {NoStop}%
\bibitem [{\citenamefont {Patton}\ \emph {et~al.}(1988)\citenamefont {Patton},
  \citenamefont {Iyer}, \citenamefont {Delage}, \citenamefont {Tiwari},\ and\
  \citenamefont {Stork}}]{Patton1988a}%
  \BibitemOpen
  \bibfield  {author} {\bibinfo {author} {\bibfnamefont {G.~L.}\ \bibnamefont
  {Patton}}, \bibinfo {author} {\bibfnamefont {S.~S.}\ \bibnamefont {Iyer}},
  \bibinfo {author} {\bibfnamefont {S.~L.}\ \bibnamefont {Delage}}, \bibinfo
  {author} {\bibfnamefont {S.}~\bibnamefont {Tiwari}},\ and\ \bibinfo {author}
  {\bibfnamefont {J.~M.}\ \bibnamefont {Stork}},\ }\bibfield  {title} {\bibinfo
  {title} {{Silicon-germanium base heterojunction bipolar transistors by
  molecular beam epitaxy}},\ }\href {https://doi.org/10.1109/55.677} {\bibfield
   {journal} {\bibinfo  {journal} {IEEE Electron Device Letters}\ }\textbf
  {\bibinfo {volume} {9}},\ \bibinfo {pages} {165} (\bibinfo {year}
  {1988})}\BibitemShut {NoStop}%
\bibitem [{\citenamefont {Gruhle}\ \emph {et~al.}(1992)\citenamefont {Gruhle},
  \citenamefont {Kibbel}, \citenamefont {Konig}, \citenamefont {Erben},\ and\
  \citenamefont {Kasper}}]{gruhle1992mbe}%
  \BibitemOpen
  \bibfield  {author} {\bibinfo {author} {\bibfnamefont {A.}~\bibnamefont
  {Gruhle}}, \bibinfo {author} {\bibfnamefont {H.}~\bibnamefont {Kibbel}},
  \bibinfo {author} {\bibfnamefont {U.}~\bibnamefont {Konig}}, \bibinfo
  {author} {\bibfnamefont {U.}~\bibnamefont {Erben}},\ and\ \bibinfo {author}
  {\bibfnamefont {E.}~\bibnamefont {Kasper}},\ }\bibfield  {title} {\bibinfo
  {title} {{MBE-grown Si/SiGe HBTs with high beta, $f_T$, and $f_{max}$ }},\
  }\href {https://doi.org/10.1109/55.145022} {\bibfield  {journal} {\bibinfo
  {journal} {IEEE electron device letters}\ }\textbf {\bibinfo {volume} {13}},\
  \bibinfo {pages} {206} (\bibinfo {year} {1992})}\BibitemShut {NoStop}%
\bibitem [{\citenamefont {Patton}\ \emph {et~al.}(1990)\citenamefont {Patton},
  \citenamefont {Comfort}, \citenamefont {Meyerson},\ and\ \citenamefont
  {Emmanuel}}]{Patton1990}%
  \BibitemOpen
  \bibfield  {author} {\bibinfo {author} {\bibfnamefont {G.~L.}\ \bibnamefont
  {Patton}}, \bibinfo {author} {\bibfnamefont {J.~H.}\ \bibnamefont {Comfort}},
  \bibinfo {author} {\bibfnamefont {B.~S.}\ \bibnamefont {Meyerson}},\ and\
  \bibinfo {author} {\bibfnamefont {F.}~\bibnamefont {Emmanuel}},\ }\bibfield
  {title} {\bibinfo {title} {{SiGe-Base Heterojunction Bipolar Transistors}},\
  }\href {https://doi.org/10.1109/IEDM.1990.237237} {\bibfield  {journal}
  {\bibinfo  {journal} {International Technical Digest on Electron Devices}\
  }\textbf {\bibinfo {volume} {1}},\ \bibinfo {pages} {171} (\bibinfo {year}
  {1990})}\BibitemShut {NoStop}%
\bibitem [{\citenamefont {Cressler}\ \emph {et~al.}(1993)\citenamefont
  {Cressler}, \citenamefont {Comfort}, \citenamefont {Crabbe}, \citenamefont
  {Patton}, \citenamefont {Stork}, \citenamefont {Sun},\ and\ \citenamefont
  {Meyerson}}]{Cressler1993}%
  \BibitemOpen
  \bibfield  {author} {\bibinfo {author} {\bibfnamefont {J.~D.}\ \bibnamefont
  {Cressler}}, \bibinfo {author} {\bibfnamefont {J.~H.}\ \bibnamefont
  {Comfort}}, \bibinfo {author} {\bibfnamefont {E.~F.}\ \bibnamefont {Crabbe}},
  \bibinfo {author} {\bibfnamefont {G.~L.}\ \bibnamefont {Patton}}, \bibinfo
  {author} {\bibfnamefont {J.~M.}\ \bibnamefont {Stork}}, \bibinfo {author}
  {\bibfnamefont {J.~Y.}\ \bibnamefont {Sun}},\ and\ \bibinfo {author}
  {\bibfnamefont {B.~S.}\ \bibnamefont {Meyerson}},\ }\bibfield  {title}
  {\bibinfo {title} {{On the Profile Design and Optimization of Epitaxial Si
  and SiGe-Base Bipolar Technology for 77 K Applications—Part I: Transistor
  DC Design Considerations}},\ }\href {https://doi.org/10.1109/16.199358}
  {\bibfield  {journal} {\bibinfo  {journal} {IEEE Transactions on Electron
  Devices}\ }\textbf {\bibinfo {volume} {40}},\ \bibinfo {pages} {525}
  (\bibinfo {year} {1993})}\BibitemShut {NoStop}%
\bibitem [{\citenamefont {Joseph}\ \emph {et~al.}(1995)\citenamefont {Joseph},
  \citenamefont {Cressler},\ and\ \citenamefont {Richey}}]{Joseph1995}%
  \BibitemOpen
  \bibfield  {author} {\bibinfo {author} {\bibfnamefont {A.~J.}\ \bibnamefont
  {Joseph}}, \bibinfo {author} {\bibfnamefont {J.~D.}\ \bibnamefont
  {Cressler}},\ and\ \bibinfo {author} {\bibfnamefont {D.~M.}\ \bibnamefont
  {Richey}},\ }\bibfield  {title} {\bibinfo {title} {{Operation of SiGe
  Heterojunction Bipolar Transistors in the Liquid-Helium Temperature
  Regime}},\ }\href {https://doi.org/10.1109/55.790731} {\bibfield  {journal}
  {\bibinfo  {journal} {IEEE Electron Device Letters}\ }\textbf {\bibinfo
  {volume} {16}},\ \bibinfo {pages} {268} (\bibinfo {year} {1995})}\BibitemShut
  {NoStop}%
\bibitem [{\citenamefont {Richey}\ \emph {et~al.}(1996)\citenamefont {Richey},
  \citenamefont {Joseph}, \citenamefont {Cressler},\ and\ \citenamefont
  {Jaeger}}]{Richey1996}%
  \BibitemOpen
  \bibfield  {author} {\bibinfo {author} {\bibfnamefont {D.~M.}\ \bibnamefont
  {Richey}}, \bibinfo {author} {\bibfnamefont {A.~J.}\ \bibnamefont {Joseph}},
  \bibinfo {author} {\bibfnamefont {J.~D.}\ \bibnamefont {Cressler}},\ and\
  \bibinfo {author} {\bibfnamefont {R.~C.}\ \bibnamefont {Jaeger}},\ }\bibfield
   {title} {\bibinfo {title} {{Evidence for non-equilibrium base transport in
  Si and SiGe bipolar transistors at cryogenic temperatures}},\ }\href
  {https://doi.org/10.1016/0038-1101(95)00223-5} {\bibfield  {journal}
  {\bibinfo  {journal} {Solid-State Electronics}\ }\textbf {\bibinfo {volume}
  {39}},\ \bibinfo {pages} {785} (\bibinfo {year} {1996})}\BibitemShut
  {NoStop}%
\bibitem [{\citenamefont {Rucker}\ \emph {et~al.}(2017)\citenamefont {Rucker},
  \citenamefont {Korn},\ and\ \citenamefont {Schmidt}}]{Rucker2017}%
  \BibitemOpen
  \bibfield  {author} {\bibinfo {author} {\bibfnamefont {H.}~\bibnamefont
  {Rucker}}, \bibinfo {author} {\bibfnamefont {J.}~\bibnamefont {Korn}},\ and\
  \bibinfo {author} {\bibfnamefont {J.}~\bibnamefont {Schmidt}},\ }\bibfield
  {title} {\bibinfo {title} {{Operation of SiGe HBTs at cryogenic
  temperatures}},\ }\href {https://doi.org/10.1109/BCTM.2017.8112902}
  {\bibfield  {journal} {\bibinfo  {journal} {Proceedings of the IEEE
  Bipolar/BiCMOS Circuits and Technology Meeting}\ ,\ \bibinfo {pages} {17}}
  (\bibinfo {year} {2017})}\BibitemShut {NoStop}%
\bibitem [{\citenamefont {Davidovi{\'{c}}}\ \emph {et~al.}(2017)\citenamefont
  {Davidovi{\'{c}}}, \citenamefont {Ying}, \citenamefont {Dark}, \citenamefont
  {Wier}, \citenamefont {Ge}, \citenamefont {Lourenco}, \citenamefont
  {Omprakash}, \citenamefont {Mourigal},\ and\ \citenamefont
  {Cressler}}]{Davidovic2017}%
  \BibitemOpen
  \bibfield  {author} {\bibinfo {author} {\bibfnamefont {D.}~\bibnamefont
  {Davidovi{\'{c}}}}, \bibinfo {author} {\bibfnamefont {H.}~\bibnamefont
  {Ying}}, \bibinfo {author} {\bibfnamefont {J.}~\bibnamefont {Dark}}, \bibinfo
  {author} {\bibfnamefont {B.~R.}\ \bibnamefont {Wier}}, \bibinfo {author}
  {\bibfnamefont {L.}~\bibnamefont {Ge}}, \bibinfo {author} {\bibfnamefont
  {N.~E.}\ \bibnamefont {Lourenco}}, \bibinfo {author} {\bibfnamefont {A.~P.}\
  \bibnamefont {Omprakash}}, \bibinfo {author} {\bibfnamefont {M.}~\bibnamefont
  {Mourigal}},\ and\ \bibinfo {author} {\bibfnamefont {J.~D.}\ \bibnamefont
  {Cressler}},\ }\bibfield  {title} {\bibinfo {title} {{Tunneling, current
  gain, and transconductance in silicon-germanium heterojunction bipolar
  transistors operating at millikelvin temperatures}},\ }\href
  {https://doi.org/10.1103/PhysRevApplied.8.024015} {\bibfield  {journal}
  {\bibinfo  {journal} {Physical Review Applied}\ }\textbf {\bibinfo {volume}
  {8}},\ \bibinfo {pages} {1} (\bibinfo {year} {2017})}\BibitemShut {NoStop}%
\bibitem [{\citenamefont {Ying}\ \emph {et~al.}(2018)\citenamefont {Ying},
  \citenamefont {Dark}, \citenamefont {Omprakash}, \citenamefont {Wier},
  \citenamefont {Ge}, \citenamefont {Raghunathan}, \citenamefont {Lourenco},
  \citenamefont {Fleetwood}, \citenamefont {Mourigal}, \citenamefont
  {Davidovic},\ and\ \citenamefont {Cressler}}]{Ying2018}%
  \BibitemOpen
  \bibfield  {author} {\bibinfo {author} {\bibfnamefont {H.}~\bibnamefont
  {Ying}}, \bibinfo {author} {\bibfnamefont {J.}~\bibnamefont {Dark}}, \bibinfo
  {author} {\bibfnamefont {A.~P.}\ \bibnamefont {Omprakash}}, \bibinfo {author}
  {\bibfnamefont {B.~R.}\ \bibnamefont {Wier}}, \bibinfo {author}
  {\bibfnamefont {L.}~\bibnamefont {Ge}}, \bibinfo {author} {\bibfnamefont
  {U.}~\bibnamefont {Raghunathan}}, \bibinfo {author} {\bibfnamefont {N.~E.}\
  \bibnamefont {Lourenco}}, \bibinfo {author} {\bibfnamefont {Z.~E.}\
  \bibnamefont {Fleetwood}}, \bibinfo {author} {\bibfnamefont {M.}~\bibnamefont
  {Mourigal}}, \bibinfo {author} {\bibfnamefont {D.}~\bibnamefont
  {Davidovic}},\ and\ \bibinfo {author} {\bibfnamefont {J.~D.}\ \bibnamefont
  {Cressler}},\ }\bibfield  {title} {\bibinfo {title} {{Collector Transport in
  SiGe HBTs Operating at Cryogenic Temperatures}},\ }\href
  {https://doi.org/10.1109/TED.2018.2854288} {\bibfield  {journal} {\bibinfo
  {journal} {IEEE Transactions on Electron Devices}\ }\textbf {\bibinfo
  {volume} {65}},\ \bibinfo {pages} {3697} (\bibinfo {year}
  {2018})}\BibitemShut {NoStop}%
\bibitem [{\citenamefont {Swirhun}\ \emph {et~al.}(1988)\citenamefont
  {Swirhun}, \citenamefont {Kane},\ and\ \citenamefont
  {Swanson}}]{Swirhun1988}%
  \BibitemOpen
  \bibfield  {author} {\bibinfo {author} {\bibfnamefont {S.~E.}\ \bibnamefont
  {Swirhun}}, \bibinfo {author} {\bibfnamefont {D.~E.}\ \bibnamefont {Kane}},\
  and\ \bibinfo {author} {\bibfnamefont {R.~M.}\ \bibnamefont {Swanson}},\
  }\bibfield  {title} {\bibinfo {title} {{Temperature dependence of minority
  electron mobility and bandgap narrowing in p+ Si}},\ }\href
  {https://doi.org/10.1109/iedm.1988.32816} {\bibfield  {journal} {\bibinfo
  {journal} {Technical Digest - International Electron Devices Meeting}\ ,\
  \bibinfo {pages} {298}} (\bibinfo {year} {1988})}\BibitemShut {NoStop}%
\bibitem [{\citenamefont {Klaassen}(1992)}]{Klaassenn1992}%
  \BibitemOpen
  \bibfield  {author} {\bibinfo {author} {\bibfnamefont {D.~B.~M.}\
  \bibnamefont {Klaassen}},\ }\bibfield  {title} {\bibinfo {title} {{A unified
  mobility model for device simulation—II. Temperature dependence of carrier
  mobility and lifetime}},\ }\href
  {https://doi.org/https://doi.org/10.1016/0038-1101(92)90326-8} {\bibfield
  {journal} {\bibinfo  {journal} {Solid State Electronics}\ }\textbf {\bibinfo
  {volume} {35}},\ \bibinfo {pages} {961} (\bibinfo {year} {1992})}\BibitemShut
  {NoStop}%
\bibitem [{\citenamefont {Rieh}(2000)}]{Rieh2000a}%
  \BibitemOpen
  \bibfield  {author} {\bibinfo {author} {\bibfnamefont {J.~S.}\ \bibnamefont
  {Rieh}},\ }\bibfield  {title} {\bibinfo {title} {{Temperature dependent
  minority electron mobilities in strained Si/sub 1-x/Ge/sub x/ (0. 2
  {\textless} x {\textless} 0. 4) layers}},\ }\href
  {https://doi.org/10.1109/16.831009} {\bibfield  {journal} {\bibinfo
  {journal} {IEEE Transactions on Electron Devices}\ }\textbf {\bibinfo
  {volume} {47}},\ \bibinfo {pages} {883} (\bibinfo {year} {2000})}\BibitemShut
  {NoStop}%
\bibitem [{\citenamefont {Lundstrom}(2000)}]{Lundstrom2000fundamentals}%
  \BibitemOpen
  \bibfield  {author} {\bibinfo {author} {\bibfnamefont {M.}~\bibnamefont
  {Lundstrom}},\ }\href {https://doi.org/10.1017/cbo9780511618611} {\emph
  {\bibinfo {title} {Fundamentals of Carrier Transport}}}\ (\bibinfo {year}
  {2000})\BibitemShut {NoStop}%
\bibitem [{\citenamefont {Chen}(2005)}]{chen2005nanoscale}%
  \BibitemOpen
  \bibfield  {author} {\bibinfo {author} {\bibfnamefont {G.}~\bibnamefont
  {Chen}},\ }\href@noop {} {\emph {\bibinfo {title} {{Nanoscale energy
  transport and conversion: a parallel treatment of electrons, molecules,
  phonons, and photons}}}}\ (\bibinfo  {publisher} {Oxford university press},\
  \bibinfo {year} {2005})\BibitemShut {NoStop}%
\bibitem [{\citenamefont {Chandrasekhar}(1960)}]{chandrasekhar1960radiative}%
  \BibitemOpen
  \bibfield  {author} {\bibinfo {author} {\bibfnamefont {S.}~\bibnamefont
  {Chandrasekhar}},\ }\href@noop {} {\emph {\bibinfo {title} {{Radiative
  transfer}}}}\ (\bibinfo  {publisher} {Dover Publ.(New York, NY)},\ \bibinfo
  {year} {1960})\BibitemShut {NoStop}%
\bibitem [{\citenamefont {Modest}(2013)}]{modest2013radiative}%
  \BibitemOpen
  \bibfield  {author} {\bibinfo {author} {\bibfnamefont {M.~F.}\ \bibnamefont
  {Modest}},\ }\href@noop {} {\emph {\bibinfo {title} {{Radiative heat
  transfer}}}}\ (\bibinfo  {publisher} {Academic press},\ \bibinfo {year}
  {2013})\BibitemShut {NoStop}%
\bibitem [{\citenamefont {Spanier}\ and\ \citenamefont
  {Gelbard}(1969)}]{spanier1969monte}%
  \BibitemOpen
  \bibfield  {author} {\bibinfo {author} {\bibfnamefont {J.}~\bibnamefont
  {Spanier}}\ and\ \bibinfo {author} {\bibfnamefont {E.}~\bibnamefont
  {Gelbard}},\ }\href {https://books.google.com/books?id=Gcd-AAAAIAAJ} {\emph
  {\bibinfo {title} {{Monte Carlo Principles and Neutron Transport
  Problems}}}},\ Addison-Wesley series in computer science and information
  processing\ (\bibinfo  {publisher} {Addison-Wesley Publishing Company},\
  \bibinfo {year} {1969})\BibitemShut {NoStop}%
\bibitem [{\citenamefont {Hua}\ and\ \citenamefont {Minnich}(2015)}]{Hua2015}%
  \BibitemOpen
  \bibfield  {author} {\bibinfo {author} {\bibfnamefont {C.}~\bibnamefont
  {Hua}}\ and\ \bibinfo {author} {\bibfnamefont {A.~J.}\ \bibnamefont
  {Minnich}},\ }\bibfield  {title} {\bibinfo {title} {{Semi-analytical solution
  to the frequency-dependent Boltzmann transport equation for cross-plane heat
  conduction in thin films}},\ }\bibfield  {journal} {\bibinfo  {journal}
  {Journal of Applied Physics}\ }\textbf {\bibinfo {volume} {117}},\ \href
  {https://doi.org/10.1063/1.4919432} {10.1063/1.4919432} (\bibinfo {year}
  {2015})\BibitemShut {NoStop}%
\bibitem [{\citenamefont {Grinberg}\ and\ \citenamefont
  {Luryi}(1992)}]{grinberg1992diffusion}%
  \BibitemOpen
  \bibfield  {author} {\bibinfo {author} {\bibfnamefont {A.~A.}\ \bibnamefont
  {Grinberg}}\ and\ \bibinfo {author} {\bibfnamefont {S.}~\bibnamefont
  {Luryi}},\ }\bibfield  {title} {\bibinfo {title} {{Diffusion in a short
  base}},\ }\href
  {https://doi.org/https://doi.org/10.1016/0038-1101(92)90165-9} {\bibfield
  {journal} {\bibinfo  {journal} {Solid-state electronics}\ }\textbf {\bibinfo
  {volume} {35}},\ \bibinfo {pages} {1299} (\bibinfo {year}
  {1992})}\BibitemShut {NoStop}%
\bibitem [{\citenamefont {Stettler}\ and\ \citenamefont
  {Lundstrom}(1994)}]{Stettler1994}%
  \BibitemOpen
  \bibfield  {author} {\bibinfo {author} {\bibfnamefont {M.~A.}\ \bibnamefont
  {Stettler}}\ and\ \bibinfo {author} {\bibfnamefont {M.~S.}\ \bibnamefont
  {Lundstrom}},\ }\bibfield  {title} {\bibinfo {title} {{A Microscopic Study of
  Transport in Thin Base Silicon Bipolar Transistors}},\ }\href
  {https://doi.org/10.1109/16.293317} {\bibfield  {journal} {\bibinfo
  {journal} {IEEE Transactions on Electron Devices}\ }\textbf {\bibinfo
  {volume} {41}},\ \bibinfo {pages} {1027} (\bibinfo {year}
  {1994})}\BibitemShut {NoStop}%
\bibitem [{\citenamefont {Tanaka}\ and\ \citenamefont
  {Lundstrom}(1994)}]{Tanaka}%
  \BibitemOpen
  \bibfield  {author} {\bibinfo {author} {\bibfnamefont {S.~I.}\ \bibnamefont
  {Tanaka}}\ and\ \bibinfo {author} {\bibfnamefont {M.}~\bibnamefont
  {Lundstrom}},\ }\bibfield  {title} {\bibinfo {title} {A compact hbt device
  model based on a one-flux treatment of carrier transport},\ }\href
  {https://doi.org/https://doi.org/10.1016/0038-1101(94)90004-3} {\bibfield
  {journal} {\bibinfo  {journal} {Solid-state electronics}\ }\textbf {\bibinfo
  {volume} {37}},\ \bibinfo {pages} {401} (\bibinfo {year} {1994})}\BibitemShut
  {NoStop}%
\bibitem [{\citenamefont {{\"O}zaydin}\ and\ \citenamefont
  {Eastman}(1996)}]{ozaydin1996non}%
  \BibitemOpen
  \bibfield  {author} {\bibinfo {author} {\bibfnamefont {M.}~\bibnamefont
  {{\"O}zaydin}}\ and\ \bibinfo {author} {\bibfnamefont {L.~F.}\ \bibnamefont
  {Eastman}},\ }\bibfield  {title} {\bibinfo {title} {{Non-equilibrium carrier
  transport in the base of heterojunction bipolar transistors}},\ }\href
  {https://doi.org/https://doi.org/10.1016/0038-1101(95)00159-X} {\bibfield
  {journal} {\bibinfo  {journal} {Solid-State Electronics}\ }\textbf {\bibinfo
  {volume} {39}},\ \bibinfo {pages} {731} (\bibinfo {year} {1996})}\BibitemShut
  {NoStop}%
\bibitem [{\citenamefont {Cercignani}\ and\ \citenamefont
  {Daneri}(1963)}]{cercignani1963flow}%
  \BibitemOpen
  \bibfield  {author} {\bibinfo {author} {\bibfnamefont {C.}~\bibnamefont
  {Cercignani}}\ and\ \bibinfo {author} {\bibfnamefont {A.}~\bibnamefont
  {Daneri}},\ }\bibfield  {title} {\bibinfo {title} {{Flow of a rarefied gas
  between two parallel plates}},\ }\href
  {https://doi.org/https://doi.org/10.1063/1.1729249} {\bibfield  {journal}
  {\bibinfo  {journal} {Journal of Applied Physics}\ }\textbf {\bibinfo
  {volume} {34}},\ \bibinfo {pages} {3509} (\bibinfo {year}
  {1963})}\BibitemShut {NoStop}%
\bibitem [{\citenamefont {Cercignani}(2000)}]{cercignani2000rarefied}%
  \BibitemOpen
  \bibfield  {author} {\bibinfo {author} {\bibfnamefont {C.}~\bibnamefont
  {Cercignani}},\ }\href@noop {} {\emph {\bibinfo {title} {{Rarefied gas
  dynamics: from basic concepts to actual calculations}}}},\ Vol.~\bibinfo
  {volume} {21}\ (\bibinfo  {publisher} {Cambridge University Press},\ \bibinfo
  {year} {2000})\BibitemShut {NoStop}%
\bibitem [{\citenamefont {Manela}\ and\ \citenamefont
  {Hadjiconstantinou}(2008)}]{Manela2008}%
  \BibitemOpen
  \bibfield  {author} {\bibinfo {author} {\bibfnamefont {A.}~\bibnamefont
  {Manela}}\ and\ \bibinfo {author} {\bibfnamefont {N.~G.}\ \bibnamefont
  {Hadjiconstantinou}},\ }\bibfield  {title} {\bibinfo {title} {{Gas motion
  induced by unsteady boundary heating in a small-scale slab}},\ }\bibfield
  {journal} {\bibinfo  {journal} {Physics of Fluids}\ }\textbf {\bibinfo
  {volume} {20}},\ \href {https://doi.org/10.1063/1.3010759}
  {10.1063/1.3010759} (\bibinfo {year} {2008})\BibitemShut {NoStop}%
\bibitem [{\citenamefont {Lundstrom}(2018)}]{Lundstrom2018}%
  \BibitemOpen
  \bibfield  {author} {\bibinfo {author} {\bibfnamefont {M.}~\bibnamefont
  {Lundstrom}},\ }\bibfield  {title} {\bibinfo {title} {{Carrier Transport in
  BJTs: From Ballistic to Diffusive and Off-Equilibrium}},\ }\href
  {https://doi.org/10.1109/BCICTS.2018.8551154} {\bibfield  {journal} {\bibinfo
   {journal} {2018 IEEE BiCMOS and Compound Semiconductor Integrated Circuits
  and Technology Symposium, BCICTS 2018}\ ,\ \bibinfo {pages} {174}} (\bibinfo
  {year} {2018})}\BibitemShut {NoStop}%
\bibitem [{\citenamefont {Grad}(1963)}]{grad1963asymptotic}%
  \BibitemOpen
  \bibfield  {author} {\bibinfo {author} {\bibfnamefont {H.}~\bibnamefont
  {Grad}},\ }\bibfield  {title} {\bibinfo {title} {{Asymptotic theory of the
  Boltzmann equation}},\ }\href
  {https://doi.org/https://doi.org/10.1063/1.1706716} {\bibfield  {journal}
  {\bibinfo  {journal} {The physics of Fluids}\ }\textbf {\bibinfo {volume}
  {6}},\ \bibinfo {pages} {147} (\bibinfo {year} {1963})}\BibitemShut {NoStop}%
\bibitem [{\citenamefont {Sone}(1969)}]{sone1969asymptotic}%
  \BibitemOpen
  \bibfield  {author} {\bibinfo {author} {\bibfnamefont {Y.}~\bibnamefont
  {Sone}},\ }\bibfield  {title} {\bibinfo {title} {{Asymptotic theory of flow
  of rarefied gas over a smooth boundary I}},\ }\href@noop {} {\bibfield
  {journal} {\bibinfo  {journal} {Rarefied Gas Dynamics}\ ,\ \bibinfo {pages}
  {243}} (\bibinfo {year} {1969})}\BibitemShut {NoStop}%
\bibitem [{\citenamefont {P{\'{e}}raud}\ and\ \citenamefont
  {Hadjiconstantinou}(2016)}]{Peraud2016a}%
  \BibitemOpen
  \bibfield  {author} {\bibinfo {author} {\bibfnamefont {J.~P.~M.}\
  \bibnamefont {P{\'{e}}raud}}\ and\ \bibinfo {author} {\bibfnamefont {N.~G.}\
  \bibnamefont {Hadjiconstantinou}},\ }\bibfield  {title} {\bibinfo {title}
  {{Extending the range of validity of Fourier's law into the kinetic transport
  regime via asymptotic solution of the phonon Boltzmann transport equation}},\
  }\bibfield  {journal} {\bibinfo  {journal} {Physical Review B}\ }\textbf
  {\bibinfo {volume} {93}},\ \href {https://doi.org/10.1103/PhysRevB.93.045424}
  {10.1103/PhysRevB.93.045424} (\bibinfo {year} {2016})\BibitemShut {NoStop}%
\bibitem [{\citenamefont {Tung}\ \emph {et~al.}(1991)\citenamefont {Tung},
  \citenamefont {Levi}, \citenamefont {Sullivan},\ and\ \citenamefont
  {Schrey}}]{Tung1991nisi}%
  \BibitemOpen
  \bibfield  {author} {\bibinfo {author} {\bibfnamefont {R.~T.}\ \bibnamefont
  {Tung}}, \bibinfo {author} {\bibfnamefont {A.~F.~J.}\ \bibnamefont {Levi}},
  \bibinfo {author} {\bibfnamefont {J.~P.}\ \bibnamefont {Sullivan}},\ and\
  \bibinfo {author} {\bibfnamefont {F.}~\bibnamefont {Schrey}},\ }\bibfield
  {title} {\bibinfo {title} {{Schottky-barrier inhomogeneity at epitaxial
  ${\mathrm{NiSi}}_{2}$ interfaces on Si(100)}},\ }\href
  {https://doi.org/10.1103/PhysRevLett.66.72} {\bibfield  {journal} {\bibinfo
  {journal} {Phys. Rev. Lett.}\ }\textbf {\bibinfo {volume} {66}},\ \bibinfo
  {pages} {72} (\bibinfo {year} {1991})}\BibitemShut {NoStop}%
\bibitem [{\citenamefont {Tung}(1992)}]{Tung1992}%
  \BibitemOpen
  \bibfield  {author} {\bibinfo {author} {\bibfnamefont {R.~T.}\ \bibnamefont
  {Tung}},\ }\bibfield  {title} {\bibinfo {title} {Electron transport at
  metal-semiconductor interfaces: General theory},\ }\href
  {https://doi.org/10.1103/PhysRevB.45.13509} {\bibfield  {journal} {\bibinfo
  {journal} {Phys. Rev. B}\ }\textbf {\bibinfo {volume} {45}},\ \bibinfo
  {pages} {13509} (\bibinfo {year} {1992})}\BibitemShut {NoStop}%
\bibitem [{\citenamefont {von Haartman}\ \emph {et~al.}(2002)\citenamefont {von
  Haartman}, \citenamefont {Sand{\'e}n}, \citenamefont {{\"O}stling},\ and\
  \citenamefont {Bosman}}]{vonHaartman2002rts}%
  \BibitemOpen
  \bibfield  {author} {\bibinfo {author} {\bibfnamefont {M.}~\bibnamefont {von
  Haartman}}, \bibinfo {author} {\bibfnamefont {M.}~\bibnamefont {Sand{\'e}n}},
  \bibinfo {author} {\bibfnamefont {M.}~\bibnamefont {{\"O}stling}},\ and\
  \bibinfo {author} {\bibfnamefont {G.}~\bibnamefont {Bosman}},\ }\bibfield
  {title} {\bibinfo {title} {{Random telegraph signal noise in SiGe
  heterojunction bipolar transistors}},\ }\href
  {https://doi.org/https://doi.org/10.1063/1.1506197} {\bibfield  {journal}
  {\bibinfo  {journal} {Journal of applied physics}\ }\textbf {\bibinfo
  {volume} {92}},\ \bibinfo {pages} {4414} (\bibinfo {year}
  {2002})}\BibitemShut {NoStop}%
\bibitem [{\citenamefont {Maassen}\ and\ \citenamefont
  {Lundstrom}(2015{\natexlab{a}})}]{Maassen2015a}%
  \BibitemOpen
  \bibfield  {author} {\bibinfo {author} {\bibfnamefont {J.}~\bibnamefont
  {Maassen}}\ and\ \bibinfo {author} {\bibfnamefont {M.}~\bibnamefont
  {Lundstrom}},\ }\bibfield  {title} {\bibinfo {title} {{Steady-state heat
  transport: Ballistic-to-diffusive with Fourier's law}},\ }\bibfield
  {journal} {\bibinfo  {journal} {Journal of Applied Physics}\ }\textbf
  {\bibinfo {volume} {117}},\ \href {https://doi.org/10.1063/1.4905590}
  {10.1063/1.4905590} (\bibinfo {year} {2015}{\natexlab{a}}),\ \Eprint
  {https://arxiv.org/abs/1408.1631} {arXiv:1408.1631} \BibitemShut {NoStop}%
\bibitem [{\citenamefont {Maassen}\ and\ \citenamefont
  {Lundstrom}(2015{\natexlab{b}})}]{Maassen2015}%
  \BibitemOpen
  \bibfield  {author} {\bibinfo {author} {\bibfnamefont {J.}~\bibnamefont
  {Maassen}}\ and\ \bibinfo {author} {\bibfnamefont {M.}~\bibnamefont
  {Lundstrom}},\ }\bibfield  {title} {\bibinfo {title} {{(Invited) The Landauer
  Approach to Electron and Phonon Transport}},\ }\href
  {https://doi.org/10.1149/06909.0023ecst} {\bibfield  {journal} {\bibinfo
  {journal} {ECS Transactions}\ }\textbf {\bibinfo {volume} {69}},\ \bibinfo
  {pages} {23} (\bibinfo {year} {2015}{\natexlab{b}})}\BibitemShut {NoStop}%
\bibitem [{iof()}]{ioffeSi}%
  \BibitemOpen
  \href {"http://www.ioffe.ru/SVA/NSM/Semicond/Si/bandstr.html"} {\bibinfo
  {title} {{NSM Archives, Si - Silicon}}},\ \bibinfo {note} {[Online; accessed
  29-April-2021]}\BibitemShut {NoStop}%
\bibitem [{\citenamefont {Choi}\ \emph {et~al.}(2021)\citenamefont {Choi},
  \citenamefont {Cheng}, \citenamefont {Hatanp{\"a}{\"a}},\ and\ \citenamefont
  {Minnich}}]{choi2021electronic}%
  \BibitemOpen
  \bibfield  {author} {\bibinfo {author} {\bibfnamefont {A.~Y.}\ \bibnamefont
  {Choi}}, \bibinfo {author} {\bibfnamefont {P.~S.}\ \bibnamefont {Cheng}},
  \bibinfo {author} {\bibfnamefont {B.}~\bibnamefont {Hatanp{\"a}{\"a}}},\ and\
  \bibinfo {author} {\bibfnamefont {A.~J.}\ \bibnamefont {Minnich}},\
  }\bibfield  {title} {\bibinfo {title} {{Electronic noise of warm electrons in
  semiconductors from first principles}},\ }\href
  {https://doi.org/https://doi.org/10.1103/PhysRevMaterials.5.044603}
  {\bibfield  {journal} {\bibinfo  {journal} {Physical Review Materials}\
  }\textbf {\bibinfo {volume} {5}},\ \bibinfo {pages} {044603} (\bibinfo {year}
  {2021})}\BibitemShut {NoStop}%
\bibitem [{\citenamefont {Schr{\"{o}}ter}\ \emph {et~al.}(2011)\citenamefont
  {Schr{\"{o}}ter}, \citenamefont {Wedel}, \citenamefont {Heinemann},
  \citenamefont {Jungemann}, \citenamefont {Krause}, \citenamefont
  {Chevalier},\ and\ \citenamefont {Chantre}}]{Schroter2011}%
  \BibitemOpen
  \bibfield  {author} {\bibinfo {author} {\bibfnamefont {M.}~\bibnamefont
  {Schr{\"{o}}ter}}, \bibinfo {author} {\bibfnamefont {G.}~\bibnamefont
  {Wedel}}, \bibinfo {author} {\bibfnamefont {B.}~\bibnamefont {Heinemann}},
  \bibinfo {author} {\bibfnamefont {C.}~\bibnamefont {Jungemann}}, \bibinfo
  {author} {\bibfnamefont {J.}~\bibnamefont {Krause}}, \bibinfo {author}
  {\bibfnamefont {P.}~\bibnamefont {Chevalier}},\ and\ \bibinfo {author}
  {\bibfnamefont {A.}~\bibnamefont {Chantre}},\ }\bibfield  {title} {\bibinfo
  {title} {{Physical and electrical performance limits of high-speed SiGeC
  HBTs- Part I: Vertical scaling}},\ }\href
  {https://doi.org/10.1109/TED.2011.2163722} {\bibfield  {journal} {\bibinfo
  {journal} {IEEE Transactions on Electron Devices}\ }\textbf {\bibinfo
  {volume} {58}},\ \bibinfo {pages} {3687} (\bibinfo {year}
  {2011})}\BibitemShut {NoStop}%
\bibitem [{\citenamefont {Jain}\ and\ \citenamefont
  {Rodwell}(2011)}]{jainRodwell2011}%
  \BibitemOpen
  \bibfield  {author} {\bibinfo {author} {\bibfnamefont {V.}~\bibnamefont
  {Jain}}\ and\ \bibinfo {author} {\bibfnamefont {M.~J.~W.}\ \bibnamefont
  {Rodwell}},\ }\bibfield  {title} {\bibinfo {title} {{Transconductance
  Degradation in Near-THz InP Double-Heterojunction Bipolar Transistors}},\
  }\href {https://doi.org/10.1109/LED.2011.2157451} {\bibfield  {journal}
  {\bibinfo  {journal} {IEEE Electron Device Letters}\ }\textbf {\bibinfo
  {volume} {32}},\ \bibinfo {pages} {1068} (\bibinfo {year}
  {2011})}\BibitemShut {NoStop}%
\bibitem [{\citenamefont {Ewing}\ \emph {et~al.}(2007)\citenamefont {Ewing},
  \citenamefont {Wahab}, \citenamefont {Ciechonski}, \citenamefont
  {Syv{\"{a}}j{\"{a}}rvi}, \citenamefont {Yakimova},\ and\ \citenamefont
  {Porter}}]{ewing2007}%
  \BibitemOpen
  \bibfield  {author} {\bibinfo {author} {\bibfnamefont {D.~J.}\ \bibnamefont
  {Ewing}}, \bibinfo {author} {\bibfnamefont {Q.}~\bibnamefont {Wahab}},
  \bibinfo {author} {\bibfnamefont {R.~R.}\ \bibnamefont {Ciechonski}},
  \bibinfo {author} {\bibfnamefont {M.}~\bibnamefont {Syv{\"{a}}j{\"{a}}rvi}},
  \bibinfo {author} {\bibfnamefont {R.}~\bibnamefont {Yakimova}},\ and\
  \bibinfo {author} {\bibfnamefont {L.~M.}\ \bibnamefont {Porter}},\ }\bibfield
   {title} {\bibinfo {title} {{Inhomogeneous electrical characteristics in
  4H-SiC Schottky diodes}},\ }\href
  {https://doi.org/10.1088/0268-1242/22/12/008} {\bibfield  {journal} {\bibinfo
   {journal} {Semiconductor Science and Technology}\ }\textbf {\bibinfo
  {volume} {22}},\ \bibinfo {pages} {1287} (\bibinfo {year}
  {2007})}\BibitemShut {NoStop}%
\bibitem [{\citenamefont {Esaki}\ \emph {et~al.}(1960)\citenamefont {Esaki},
  \citenamefont {Miyahara},\ and\ \citenamefont {Corporation}}]{esaki1960}%
  \BibitemOpen
  \bibfield  {author} {\bibinfo {author} {\bibfnamefont {L.}~\bibnamefont
  {Esaki}}, \bibinfo {author} {\bibfnamefont {Y.}~\bibnamefont {Miyahara}},\
  and\ \bibinfo {author} {\bibfnamefont {S.}~\bibnamefont {Corporation}},\
  }\bibfield  {title} {\bibinfo {title} {{A new device using the tunneling
  process in narrow p-n junctions}},\ }\href
  {https://doi.org/10.1016/0038-1101(60)90052-6} {\bibfield  {journal}
  {\bibinfo  {journal} {Solid State Electronics}\ }\textbf {\bibinfo {volume}
  {1}},\ \bibinfo {pages} {13} (\bibinfo {year} {1960})}\BibitemShut {NoStop}%
\bibitem [{\citenamefont {Dumin}\ and\ \citenamefont
  {Pearson}(1965)}]{Dumin1965}%
  \BibitemOpen
  \bibfield  {author} {\bibinfo {author} {\bibfnamefont {D.~J.}\ \bibnamefont
  {Dumin}}\ and\ \bibinfo {author} {\bibfnamefont {G.~L.}\ \bibnamefont
  {Pearson}},\ }\bibfield  {title} {\bibinfo {title} {{Properties of gallium
  arsenide diodes between 4.2°and 300°K}},\ }\href
  {https://doi.org/10.1063/1.1703009} {\bibfield  {journal} {\bibinfo
  {journal} {Journal of Applied Physics}\ }\textbf {\bibinfo {volume} {36}},\
  \bibinfo {pages} {3418} (\bibinfo {year} {1965})}\BibitemShut {NoStop}%
\bibitem [{\citenamefont {{Del Alamo}}\ and\ \citenamefont
  {Swanson}(1986)}]{delAlamo1986}%
  \BibitemOpen
  \bibfield  {author} {\bibinfo {author} {\bibfnamefont {J.~A.}\ \bibnamefont
  {{Del Alamo}}}\ and\ \bibinfo {author} {\bibfnamefont {R.~M.}\ \bibnamefont
  {Swanson}},\ }\bibfield  {title} {\bibinfo {title} {{Forward-bias tunneling:
  a limitation to bipolar device scaling}},\ }\href
  {https://doi.org/10.1109/edl.1986.26499} {\bibfield  {journal} {\bibinfo
  {journal} {Electron device letters}\ }\textbf {\bibinfo {volume} {EDL-7}},\
  \bibinfo {pages} {629} (\bibinfo {year} {1986})}\BibitemShut {NoStop}%
\bibitem [{\citenamefont {Padovani}\ and\ \citenamefont
  {Sumner}(1965)}]{padovani1965experimental}%
  \BibitemOpen
  \bibfield  {author} {\bibinfo {author} {\bibfnamefont {F.}~\bibnamefont
  {Padovani}}\ and\ \bibinfo {author} {\bibfnamefont {G.}~\bibnamefont
  {Sumner}},\ }\bibfield  {title} {\bibinfo {title} {Experimental study of
  gold-gallium arsenide schottky barriers},\ }\href
  {https://doi.org/https://doi.org/10.1063/1.1713940} {\bibfield  {journal}
  {\bibinfo  {journal} {Journal of Applied Physics}\ }\textbf {\bibinfo
  {volume} {36}},\ \bibinfo {pages} {3744} (\bibinfo {year}
  {1965})}\BibitemShut {NoStop}%
\bibitem [{\citenamefont {Padovani}\ and\ \citenamefont
  {Stratton}(1966)}]{padovani1966field}%
  \BibitemOpen
  \bibfield  {author} {\bibinfo {author} {\bibfnamefont {F.}~\bibnamefont
  {Padovani}}\ and\ \bibinfo {author} {\bibfnamefont {R.}~\bibnamefont
  {Stratton}},\ }\bibfield  {title} {\bibinfo {title} {Field and
  thermionic-field emission in schottky barriers},\ }\href
  {https://doi.org/https://doi.org/10.1016/0038-1101(66)90097-9} {\bibfield
  {journal} {\bibinfo  {journal} {Solid-State Electronics}\ }\textbf {\bibinfo
  {volume} {9}},\ \bibinfo {pages} {695} (\bibinfo {year} {1966})}\BibitemShut
  {NoStop}%
\bibitem [{\citenamefont {Hackam}\ and\ \citenamefont
  {Harrop}(1972)}]{hackam1972electrical}%
  \BibitemOpen
  \bibfield  {author} {\bibinfo {author} {\bibfnamefont {R.}~\bibnamefont
  {Hackam}}\ and\ \bibinfo {author} {\bibfnamefont {P.}~\bibnamefont
  {Harrop}},\ }\bibfield  {title} {\bibinfo {title} {Electrical properties of
  nickel-low-doped n-type gallium arsenide schottky-barrier diodes},\ }\href
  {https://doi.org/10.1109/T-ED.1972.17586} {\bibfield  {journal} {\bibinfo
  {journal} {IEEE Transactions on Electron Devices}\ }\textbf {\bibinfo
  {volume} {19}},\ \bibinfo {pages} {1231} (\bibinfo {year}
  {1972})}\BibitemShut {NoStop}%
\bibitem [{\citenamefont {Bhuiyan}\ \emph {et~al.}(1988)\citenamefont
  {Bhuiyan}, \citenamefont {Martinez},\ and\ \citenamefont
  {Esteve}}]{Bhuiyan1988}%
  \BibitemOpen
  \bibfield  {author} {\bibinfo {author} {\bibfnamefont {A.~S.}\ \bibnamefont
  {Bhuiyan}}, \bibinfo {author} {\bibfnamefont {A.}~\bibnamefont {Martinez}},\
  and\ \bibinfo {author} {\bibfnamefont {D.}~\bibnamefont {Esteve}},\
  }\bibfield  {title} {\bibinfo {title} {{A new Richardson plot for non-ideal
  schottky diodes}},\ }\href {https://doi.org/10.1016/0040-6090(88)90239-8}
  {\bibfield  {journal} {\bibinfo  {journal} {Electronics}\ }\textbf {\bibinfo
  {volume} {161}},\ \bibinfo {pages} {93} (\bibinfo {year} {1988})}\BibitemShut
  {NoStop}%
\bibitem [{\citenamefont {Tung}(1984)}]{Tung1984Nisi}%
  \BibitemOpen
  \bibfield  {author} {\bibinfo {author} {\bibfnamefont {R.~T.}\ \bibnamefont
  {Tung}},\ }\bibfield  {title} {\bibinfo {title} {{Schottky-Barrier Formation
  at Single-Crystal Metal-Semiconductor Interfaces}},\ }\href
  {https://doi.org/10.1103/PhysRevLett.52.461} {\bibfield  {journal} {\bibinfo
  {journal} {Phys. Rev. Lett.}\ }\textbf {\bibinfo {volume} {52}},\ \bibinfo
  {pages} {461} (\bibinfo {year} {1984})}\BibitemShut {NoStop}%
\bibitem [{\citenamefont {Raoult}\ \emph {et~al.}(2008)\citenamefont {Raoult},
  \citenamefont {Pascal}, \citenamefont {Delseny}, \citenamefont {Marin},\ and\
  \citenamefont {Deen}}]{raoult2008}%
  \BibitemOpen
  \bibfield  {author} {\bibinfo {author} {\bibfnamefont {J.}~\bibnamefont
  {Raoult}}, \bibinfo {author} {\bibfnamefont {F.}~\bibnamefont {Pascal}},
  \bibinfo {author} {\bibfnamefont {C.}~\bibnamefont {Delseny}}, \bibinfo
  {author} {\bibfnamefont {M.}~\bibnamefont {Marin}},\ and\ \bibinfo {author}
  {\bibfnamefont {M.~J.}\ \bibnamefont {Deen}},\ }\bibfield  {title} {\bibinfo
  {title} {{Impact of carbon concentration on 1f noise and random telegraph
  signal noise in SiGe:C heterojunction bipolar transistors}},\ }\bibfield
  {journal} {\bibinfo  {journal} {Journal of Applied Physics}\ }\textbf
  {\bibinfo {volume} {103}},\ \href {https://doi.org/10.1063/1.2939252}
  {10.1063/1.2939252} (\bibinfo {year} {2008})\BibitemShut {NoStop}%
\end{thebibliography}%

\end{document}